\documentclass[%
reprint,
showkeys,
showpacs,
nofootinbib,
amsmath,amssymb,
10pt]{revtex4-1}
\usepackage{graphicx}
\usepackage{dcolumn}
\usepackage{color,soul}
\usepackage{bm}
\usepackage[colorlinks,citecolor=blue,urlcolor=blue,linkcolor=blue]{hyperref}

\usepackage[figtopcap]{subfigure}

\begin{document}

\title{Reconstruction of $f(T)$-gravity in the absence of matter}
\author{W. El Hanafy$^{1,3}$}
\email{waleed.elhanafy@bue.edu.eg}
\author{G.L. Nashed$^{1,2,3}$}%
\email{nashed@bue.edu.eg}
\affiliation{$^{1}$Centre for theoretical physics, the British University in Egypt, 11837 - P.O. Box 43, Egypt.}
\affiliation{$^{2}$Mathematics Department, Faculty of Science, Ain Shams University, Cairo, Egypt.}
\affiliation{$^{3}$Egyptian Relativity Group (ERG).}
\begin{abstract}
We derive an exact $f(T)$ gravity in the absence of ordinary matter in Friedmann-Robertson-Walker (FRW) universe, where $T$ is the teleparallel torsion scalar. We show that vanishing of the energy-momentum tensor $\mathcal{T}^{\mu \nu}$ of matter does not imply vanishing of the teleparallel torsion scalar, in contrast to general relativity, where the Ricci scalar vanishes. The theory provides an exponential (\textit{inflationary}) scale factor independent of the choice of the sectional curvature. In addition, the obtained $f(T)$ acts just like cosmological constant in the flat space model. Nevertheless, it is dynamical in non-flat models. In particular, the open universe provides a decaying pattern of the $f(T)$ contributing directly to solve the fine-tuning problem of the cosmological constant. The equation of state (EoS) of the torsion vacuum fluid has been studied in positive and negative Hubble regimes. We study the case when the torsion is made of a scalar field (\textit{tlaplon}) which acts as torsion potential. This treatment enables to induce a tlaplon field sensitive to the symmetry of the spacetime in addition to the reconstruction of its effective potential from the $f(T)$ theory. The theory provides six different versions of inflationary models. The real solutions are mainly quadratic, the complex solutions, remarkably, provide Higgs-like potential.
\end{abstract}

\maketitle
\section{Introduction}\label{S1}
Gravity is one of the most ancient powers humans' experience. However, its nature looks different from the other fundamental interactions. In modern science, gravity is one of the fundamental interactions in nature. On the cosmological scale, gravity as a long range interaction is dominant. Unlike other interactions it shows an attraction behaviour only. But recent astrophysical and cosmological observations indicate an accelerated cosmic expansion. How does an attraction force of gravity leads the universe to an accelerated expansion? On the other hand, gravitational field is a manifestation of a matter source. Again, how does an empty universe model of de Sitter (gravitational treatment) provide an accelerated expanding universe? All of these quests tell that gravity may have a repulsive side as well! What is the source, then, of this repulsion? The general relativity (GR) theory produces gravity as a curvature in a spacetime. This representation of gravity, however, is successful on the astrophysical scale, it fails to predict the accelerated cosmic expansion which may lead finally to introduce a cosmological constant to adapt GR to the cosmological observations. Introducing a cosmological constant provides a source to produce the required repulsion term in Einstein field equations, nevertheless, it leads to one of the major problems in physics called \textit{the cosmological constant problem}. Does the cosmological constant represent a real problem? or it just alarms us that we need to reexamine our knowledge of gravity. If GR formulation is restricted by the structure of Riemannian geometry, do other geometries provide us with different scopes to reexamine gravity?

Teleparallel geometry has provided a new tool to examine gravity. Within this geometry a theory of gravity equivalent to GR, the teleparallel theory of general relativity (TEGR), has been formulated \cite{M2002}. The theory introduced a new invariant (teleparallel torsion scalar) constructed from the torsion tensor instead of Ricci invariant in the Einstein-Hilbert action. Although, the two theories are equivalent in their field equations level, they have different qualities in their Lagrangian level. Unlike Ricci scalar the torsion scalar is not invariant under local Lorentz transformation (LLT) \cite{1010.1041,1012.4039}. The Born-Infeld-modified teleparallel gravity is one of the attractive variants of TEGR. The theory predicts the early cosmic inflation without an inflaton field \cite{FF07,FF08} which reflects a repulsive side of gravity. Another interesting extension of TEGR is the $f(T)$ gravity theories, this modification is by replacing the torsion scalar $T$ by an arbitrary function $f(T)$ just similar to the $f(R)$ extension of Einstein-Hilbert action. In a particular class of these theories, also, late cosmic acceleration could be accounted for with no need to dark energy (DE) \cite{BF09,L10,1008.4036,1011.0508}. Although the equivalence of the field equations between GR and TEGR, we lose this equivalence in $f(R)$ and $f(T)$ generalizations. In addition to the lake of the invariance of the $f(T)$ field equations under LLT, the $f(T)$ theories are not conformally equivalent to TEGR plus a scalar field unlike the $f(R)$ theories which are conformally equivalent to Einstein-Hilbert action plus a scalar field \cite{Y2011}.

The $f(T)$ theories have paid more attention in the last decade to exploring different aspects in astrophysics \cite{CCDDS11,FF011,FF11,IS12,CGS13,Nashed:1996,Nashed:1997,Nashed:2002,Nashed:2006,Nashed:2007,Nashed:2012,
Nashed3,Nashed5,RHTMM13,BFG15,IRR2015,KAAS13} and in cosmology \cite{1205.3421,BNO14,BO14,JMM14,HLOS14,NH14,WCWW2015,1503.07427,HN15,RAK2015,JP2016}. Some applications show interesting results, e.g. presenting bouncing solutions avoiding trans-Planckian problems \cite{CCDDS11,CQSW14,bounce3,BNEI:2016}. A recent promising variant is the modified teleparallel equivalent of Gauss-Bonnet gravity and its applications \cite{KS114,KS214,KS314}.

In this article, we reconsider the empty universe case when it is governed by $f(T)$ gravity. This is similar to de Sitter universe which is devoid of matter energy containing only vacuum energy. Although, de Sitter solution was considered as an artificial case and has no physical significance, it becomes a foundation of many models in modern cosmology. As is well known the vacuum solutions do not require the teleparallel torsion to vanish. This may give a new insight on gravity affording a forward step to understand what is missing. The work is organized as follows: In Section \ref{S2}, we apply the general relativistic field equations to an empty FRW universe showing the need for a cosmological constant to afford a repulsive force in Einstein's field equations. In Section \ref{S3}, we review the teleparallel gravity and its extension to $f(T)$ gravity theories. In Section \ref{S4}, we apply the $f(T)$ field equations to the FRW universe in the absence of matter, obtaining an exact solution of an exponential scale factor, independent of the value of the sectional curvature $k$, in addition to exact $f(T)$ gravity. The theory is consistent with the inflationary universe. We show that the torsion density explains perfectly the missing ingredient of the vacuum. In Section \ref{S5}, we show that the decaying behavior of the obtained $f(T)$ in the open universe model contributes to solve the cosmological constant problem. In Section \ref{S6}, we investigate the EoS of the effective torsion (vacuum) gravity. Moreover, we apply a special treatment by considering the case when the torsion potential is made of a tlaplon field. This treatment enabled us to construct a scalar field sensitive to the vierbein (spacetime symmetry) with potential constructed from the adopted $f(T)$ gravity theory. In Section \ref{S7}, we reformulate the obtained $f(T)$ theory in terms of a scalar field in spatially non-flat models. Final remarks are given in Section \ref{S8}.
\section{de Sitter cosmology}\label{S2}
The initial motivation of using the cosmological constant term by Einstein is to provide a static universe filled with matter. In 1922, Friedmann has shown that imposing the cosmological constant in the general relativity (GR) field equations does not provide a stable static universe. Later in 1929, Friedmann's result has been supported by Edwin Hubble observations which clarify the expanding pattern of the universe. On the other hand, de Sitter showed that there is another solution to Einstein's field equations corresponds to an empty universe. Eddington referred to de Sitter's model as motion without matter in contrast to Einstein's model which has matter without motion. Although, Einstein has omitted the cosmological constant referring to it as his biggest blunder, the later astronomical observations of high-redshift Type Ia supernovae calling the cosmological constant back to life \cite{SN98}. However, de Sitter model was not easy to explained at that time, his solution had been considered as a toy model with no physical significance at that time, now it becomes a foundation for many cosmological models. We show here a motivation of using the cosmological constant as a vacuum density in the GR to explain the accelerated cosmic expansion. The FRW metric of an isotropic and homogeneous universe can be written in spherical polar coordinates ($t$, $r$, $\theta$, $\phi$) as
\begin{equation}\label{FRW_metric}
    ds^2=dt^{2}-a^{2}(t) \left[\frac{dr^{2}}{1-k r^{2}}+r^{2} d\theta^2+r^{2}\sin^{2}(\theta) d\phi^{2}\right],
\end{equation}
where $a(t)$ is the scale factor. To setup our notations: we denote the reduced Planck mass as $M_{p}$, the Newton gravitational constant as $G$, the speed of light in vacuum as $c$ and the reduces Planck constant as $\hbar$. Where $M_{p}$ is related to $G$ by the relation $M_{p}=\sqrt{\hbar c/8\pi G}$. We assume the natural  units in which $G = c = \hbar = 1$. Einstein's field equations are
\begin{equation}\label{Ein_FE}
    G_{\mu \nu}:=R_{\mu\nu}-\frac{1}{2}g_{\mu \nu} R=-8 \pi \mathcal{T}_{\mu \nu},
\end{equation}
where $G_{\mu \nu}$ is Einstein tensor, $R_{\mu\nu}$ is Ricci tensor, $R$ is Ricci scalar and $\mathcal{T}_{\mu \nu}:=diag(\rho_{M},-p_{M},-p_{M},-p_{M})$ is the energy-material tensor. So the Friedmann dynamical equations are
\begin{eqnarray}
  3\left(\frac{\dot{a}}{a}\right)^2=3H^2  &=& 8\pi \rho_{M} - 3\frac{k}{a^2},\label{FRW1} \\
  3\left(\frac{\ddot{a}}{a}\right) =3 q H^2 &=& -4 \pi \left(\rho_{M}+3 p_{M}\right), \label{FRW2}
\end{eqnarray}
where the dot expresses the differentiation with respect to time, $q(=-\frac{a\ddot{a}}{\dot{a}^2})$ is the \textit{deceleration} parameter and $H(=\frac{\dot{a}}{a})$ is the \textit{Hubble} parameter. We take $\rho=\rho_{c}$, where $\rho_{c}$ is the critical density of the universe when it is full of matter and spatially flat ($k=0$), then $\rho_{c}=\frac{3H^2}{8\pi}$. Equation (\ref{FRW1}) can be rewritten as
\begin{equation}
  \Omega_{M}+\Omega_{k}=1,\label{FRW3}
\end{equation}
where $\Omega_{M}=\frac{\rho_{M}}{\rho_{c}}=\frac{\rho_{M}}{3H^2/8\pi}$ represents the \textit{matter density} parameter and $\Omega_{k}(=\frac{-k}{a^2 H^2})$ is the \textit{curvature} energy density parameter. In the GR, Ricci scalar field plays the main role to describe the gravity. We consider now an empty universe whereas $\rho_{M}=0$ and $p_{M}=0$, one can easily find that the vanishing of the energy-momentum tensor ($\mathcal{T}_{\mu \nu}$ $=$ $0$) implies the vanishing of Ricci tensor and Ricci scalar ($R$ $=$ $0$) as well.

Unfortunately, this does not provide us with information about the vacuum density; so we finally have to accept introducing of the cosmological constant $\Lambda$ in (\ref{Ein_FE}). This can be achieved by the replacement $\rho_{M}\rightarrow \rho_{M}+\rho_{\Lambda}$ and $p_{M}\rightarrow p_{M}+p_{\Lambda}$, where $\rho_{\Lambda}$ and $p_{\Lambda}$ are the cosmological constant density and pressure \cite{Liddle:2003}. This gives rise to de Sitter solution of an exponentially expanding universe $a(t)\propto e^{Ht}$ in which the Hubble constant is related to the cosmological constant $H=\sqrt{\Lambda/3}$. It is easy to show that the cosmological constant produces the negative gravitational pressure powering the accelerated expansion of the empty universe.

Since, de Sitter universe is devoid of matter energy containing only vacuum energy, the cosmological constant in de Sitter universe is sometimes thought of as the energy density of empty space. In quantum physics this can be thought of as a type of zero-point-energy. However, observations suggest that $\rho_{\Lambda}\sim 0.7 \rho_{c}$ putting an upper limit of the cosmological constant to be $\sim 10^{-35} s^{-2}$ \cite{RJ2015}. Nevertheless, the zero-point energy calculations of the quantum physics predict much larger value exceeds the upper limit by $\sim 120$ orders of magnitude. This discrepancy is known as the cosmological constant problem. Although this problem is considered as the worst prediction in the history of the theoretical physics, many authors have tried vastly to solve this problem in vain. A new gravitation theory is possibly one of the right tracks to solve this problem.
\section{Teleparallelism as an alternative gravity description}\label{S3}
Actually, Riemannian and teleparallel geometries might be considered as two extremes of Riemann-Cartan geometry. The first is powered by a torsionless connection (Levi-Civita), the second is powered by curvatureless connection (Weitzenb\"ock). However, the double contraction of the first Bianchi identity of the teleparallel geometry can be delivered in the following form
\begin{equation}
\nonumber    R \equiv -T-2\nabla_{\alpha}T_{\nu}{^{\alpha\nu}},
\end{equation}
where $T_{\sigma}{^{\mu\nu}}$ is the torsion tensor of type (1,2), $T$ is called the teleparallel torsion scalar and $\nabla$ is the covariant derivative associated to the symmetric affine (Levi-Civita) connection. The integral of the total derivative $\nabla_{\alpha}T_{\nu}{^{\alpha\nu}}$ over the whole space is a boundary term so that its contribution might be discarded when the Lagrangian formalism is adopted. Accordingly, replacing Ricci scalar $R$ by the teleparallel torsion scalar $T$ in Einstein-Hilbert action provides a teleparallel equivalent theory of the general relativity (TEGR) \cite{M2002}. Similar to $f(R)$ theories, the TEGR has been generalized to $f(T)$ gravity theories \cite{BF09, L10,1008.4036,1011.0508}. In what follows we give a brief description of the teleparallel geometry.
\subsection{Installing Weitzenb\"ock connection}\label{S3S1}
The formulation of GR within the Riemannian geometry is powered by the Levi-Civita connection which gives only the attraction gravity. Nevertheless, other geometries with different qualities may give the other one, the repulsive side \cite{W2012, AP2013}. It is known that the Levi-Civita connection plays the role of the ``displacement field" in the GR, so we expect different qualities when using the Weitzenb\"ock connection of the teleparallel geometry. This space is described by as a pair $(M,~h_{i})$, where $M$ is an $n$-dimensional smooth manifold and $h_{i}$ ($i=1,\cdots, n$) are $n$ independent vector fields defined globally on $M$. The vector fields $h_{i}$ are called the parallelization vector (vielbein) fields. In the four dimensional manifold the parallelization vector fields are called the tetrad (vierbein) field. They are characterized by
\begin{equation}\label{q1}
D_{\nu} {h_i}^\mu=\partial_{\nu}
{h_i}^\mu+{\Gamma^\mu}_{\lambda \nu} {h_i}^\lambda=0,
\end{equation}
where $\partial_{\nu}:=\frac{\partial}{\partial x^{\nu}}$ and $D_{\nu}$ is the covariant derivative associated to the nonsymmetric affine (Weitzenb\"{o}ck) connection \cite{Wr}
\begin{equation}\label{q2}
{\Gamma^\lambda}_{\mu \nu} :=
{h_i}^\lambda~ \partial_\nu h^{i}{_{\mu}}.
\end{equation}
However, the vielbein space admits natural connections other than Weitzenb\"{o}ck, the later uniquely fulfills the teleparallelism condition (\ref{q1}). It is worth to see different approaches to study gravity by examining connections other than Weitzenb\"{o}ck \cite{MW77,M78,HS79,NS07,NS13,W09, WK11,YST12}, which may have an interesting astrophysical and cosmological applications \cite{Wanas86,WA13,WH14,WOR15}. One can show that equation (\ref{q1}) implies the metricity condition. Also, the curvature tensor of the connection (\ref{q2}) vanishes identically. The metric tensor $g_{\mu \nu}$ is defined by
\begin{equation}\label{q3}
 g_{\mu \nu} :=  \eta_{i j} {h^i}_\mu {h^j}_\nu,
\end{equation}
where $\eta_{i j}=(+,-,-,-)$ is the metric of Minkowski spacetime. We note that, the tetrad field ${h_i}^\mu$ determines a unique metric $g_{\mu \nu}$, while the inverse is incorrect. The torsion $T$ and the contortion $K$ tensor fields are
\begin{eqnarray}
\nonumber {T^\alpha}_{\mu \nu}  & := &
{\Gamma^\alpha}_{\nu \mu}-{\Gamma^\alpha}_{\mu \nu} ={h_i}^\alpha
\left(\partial_\mu{h^i}_\nu-\partial_\nu{h^i}_\mu\right),\\
{K^{\mu \nu}}_\alpha  & := &
-\frac{1}{2}\left({T^{\mu \nu}}_\alpha-{T^{\nu
\mu}}_\alpha-{T_\alpha}^{\mu \nu}\right). \label{q4}
\end{eqnarray}
We next define the torsion scalar of the TEGR as
\begin{equation}\label{Tor_sc}
T := {T^\alpha}_{\mu \nu}
{S_\alpha}^{\mu \nu},
\end{equation}
where the tensor ${S_\alpha}^{\mu \nu}$ is defined as
\begin{equation}\label{q5}
{S_\alpha}^{\mu \nu}
:= \frac{1}{2}\left({K^{\mu
\nu}}_\alpha+\delta^\mu_\alpha{T^{\beta
\nu}}_\beta-\delta^\nu_\alpha{T^{\beta \mu}}_\beta\right),
\end{equation}
which is skew symmetric in the last two indices. In order to get the physical meaning of the connection coefficients as the displacement field, we use (\ref{q4}) to reexpress the Weitzenb\"{o}ck connection (\ref{q2}) as
\begin{equation}\label{contortion}
    {\Gamma^\mu}_{\nu \rho }=\left \{_{\nu  \rho}^\mu\right\}+{K^{\mu}}_{\nu \rho}.
\end{equation}
By careful looking at the above expression of the new displacement field, it consists of two terms. The first is the Levi-Civita connection which consists of the gravitational potential (metric coefficients, $g_{\mu \nu}$) and its first derivatives with respect to the coordinates, while the second term is the contortion which consists of the tetrad vector fields and its first derivatives with respect to the coordinates. In this sense, we find that the first term contributes to the displacement field as the usual attractive force of gravity, while the second term contributes as a repulsive force. Now we can see how teleparallel geometry adds a new quality (torsion or contortion) to the spacetime allowing repulsive side of gravity to showup.
\subsection{$f(T)$ field equations}\label{S3S2}
Similar to the $f(R)$ theory one can define the action of $f(T)$ theory as
\begin{equation}\label{q7}
{\cal L}({h^i}_\mu, \Phi_A)=\int d^{4}x~ h\left[\frac{M_{p}^2}{2}f(T)+{\cal L}_{Matter}(\Phi_A)\right],
\end{equation}
where $ h=\sqrt{-g}=\det\left({h^a}_\mu\right)$, $\Phi_A$ are the matter fields. The variation of (\ref{q7}) with respect to the field ${h^i}_\mu$ requires the following field equations \cite{BF09}
\begin{eqnarray}\label{q8}
\nonumber &&{S_\mu}^{\rho \nu} \partial_{\rho} T f_{TT}+\left[h^{-1}{h^i}_\mu\partial_\rho\left(h{h_i}^\alpha
{S_\alpha}^{\rho \nu}\right)-{T^\alpha}_{\lambda \mu}{S_\alpha}^{\nu \lambda}\right]f_T\\
&&-\frac{1}{4}\delta^\nu_\mu f=-4\pi{{\cal T}_{\mu}}^{\nu},
\end{eqnarray}
where $f \equiv f(T)$, $f_{T}=\frac{\partial f(T)}{\partial T}$, $f_{TT}=\frac{\partial^2 f(T)}{\partial T^2}$. Although the quantitative equivalence between TEGR and GR, they are qualitatively not equivalent. For example, the Ricci scalar is invariant under local Lorentz transformation, while the total derivative term is not, so the torsion scalar. Accordingly, the $f(T)$ theories are not invariant under local Lorentz transformation \cite{1010.1041,1012.4039}. On the other hand, it is will known that $f(R)$ theories are conformally equivalent to Einstein-Hilbert action plus a scalar field. In contrast, the $f(T)$ theory cannot be conformally equivalent to TEGR plus scalar field \cite{Y2011}.
\section{$f(T)$ cosmology}\label{S4}
We apply the $f(T)$ field equations (\ref{q8}) to the FRW universe of a spatially homogeneous and isotropic spacetime, which can be described by the tetrad \cite{R32}. In spherical polar coordinates ($t$, $r$, $\theta$, $\phi$) the tetrad can be written as follows:
\begin{eqnarray}\label{tetrad}
\nonumber \left({h_{i}}^{\mu}\right)=\hspace{8cm}&\\
\nonumber \left(
\begin{tiny}  \begin{array}{cccc}
    1 & 0 & 0 & 0 \\
    0&\frac{\alpha \sin{\theta} \cos{\phi}}{4a(t)} & \frac{\beta \cos{\theta} \cos{\phi}-4r\sqrt{k}\sin{\phi}}{4 r a(t)} & -\frac{\beta \sin{\phi}+4 r \sqrt{k} \cos{\theta} \cos{\phi}}{4 r a(t)\sin{\theta}} \\[5pt]
    0&\frac{\alpha \sin{\theta} \sin{\phi}}{4 a(t)} & \frac{\beta \cos{\theta} \sin{\phi}+4 r \sqrt{k}\cos{\phi}}{4 r a(t)} & \frac{\beta \cos{\phi}-4 r \sqrt{k} \cos{\theta} \sin{\phi}}{4 r a(t)\sin{\theta}} \\[5pt]
    0&\frac{\alpha \cos{\theta}}{4 a(t)} & \frac{-\beta \sin{\theta}}{4 r a(t)} & \frac{\sqrt{k}}{a(t)} \\[5pt]
  \end{array}\end{tiny}
\right),&\\
\end{eqnarray}
where $\alpha=4+k r^{2}$ and $\beta=4-k r^{2}$. We take the case of an isotropic fluid such that the energy-momentum tensor is ${{\cal T}_{\mu}}^{\nu}=\textmd{diag}(\rho_{M},-p_{M},-p_{M},-p_{M})$. The tetrad (\ref{tetrad}) has the same metric as FRW metric (\ref{FRW_metric}). Applying the field equations (\ref{q8}) to the tetrad fields (\ref{tetrad}) gives
\begin{eqnarray}
 \mathcal{T}_{0}^{~0}\equiv\rho_{M}&=&\frac{1}{16 \pi}(f+12 H^2 f_{T}),\label{dens1}\\
\nonumber \mathcal{T}_{(a)}^{~(a)}\equiv p_{M}&=&-\frac{1}{16 \pi}\left[(f+12 H^2 f_{T}) \right.\\
\nonumber &+& 4\dot{H}(f_{T}-12 H^2 f_{TT})\\
&-&\left.\frac{4k}{a^2}(f_{T}+12H^2 f_{TT})\right],\label{press1}
\end{eqnarray}
where the index $(a)=1,2,3$, while $\rho_{M}$ and $p_{M}$ are the proper density and pressure of the matter. In order to introduce the torsion contribution to the density and pressure in the Friedmann dynamical equations we replace $\rho_{M} \rightarrow \rho_{M}+\rho_{T}$ and $p_{M} \rightarrow p_{M}+p_{T}$. Thus the $f(T)$ field equations (\ref{q8}) can be rewritten in terms of the Friedmann equations similar to GR as
\begin{eqnarray}
     3H^2    &\equiv& 8\pi\rho_{c}- 3\frac{k}{a^2}=8\pi (\rho_{M}+\rho_{T}) - 3\frac{k}{a^2},\label{TFRW1} \\
\nonumber
    3 q H^2 &\equiv& -4\pi[\rho_{c}+3p_{tot}]\\
   &=& -4 \pi \left[(\rho_{M}+\rho_{T})+3 (p_{M}+p_{T})\right], \label{TFRW2}
\end{eqnarray}
where $\rho_{c}=\rho_{M}+\rho_{T}$ and $p_{tot}=p_{M}+p_{T}$ are the critical density and the total pressure of the universe. The above equations are the modified Friedmann equations in the FRW universe governed by $f(T)$ gravity. The torsion contributes to the energy density and pressure as \cite{NH14}
\begin{eqnarray}
\rho_{T}&=&\frac{1}{8 \pi}\left(3H^2-f/2-6H^2 f_{T}+\frac{3k}{a^2}\right),\label{Tor_dens}\\
\nonumber p_{T}&=&\frac{-1}{8 \pi}\left[\frac{k}{a^2}(1+2f_{T}+24H^2 f_{TT})+2\dot{H}+3H^2\right.\\
               &-&\left.f/2-2(\dot{H}+3H^2)f_{T}+24\dot{H}H^2 f_{TT}\right].\label{Tor_press}
\end{eqnarray}
From equations (\ref{dens1}), (\ref{press1}), (\ref{Tor_dens}) and (\ref{Tor_press}) one can show that the Friedmann equations (\ref{FRW1}) and (\ref{FRW2}) are invariant under the transformation $\rho_{M} \rightarrow \rho_{M}+\rho_{T}$ and $p_{M} \rightarrow p_{M}+p_{T}$. We, also, note that in non-flat cases $T=-6H^2+6k/a^2$, this implies the inclusion of a curvature term in $\rho_{T}$ to allow its vanishing to match the TEGR limit. In more details, we find that equations (\ref{dens1})-(\ref{Tor_press}) can perform the following cases: (i) The case of $f(T)=T$, the torsion contribution ($\rho_{T}$, $p_{T}$) vanishes, while $\rho_{M}$ and $p_{M}$ reduce to the Friedmann equations of GR. (ii) The case of ($k=0$, $f(T)\neq T$), it reduces to \cite{M11}. (iii) The case of ($k\neq 0$, $f(T)\neq T$), this case shows the dynamical evolution in non-flat models \cite{NH14}. We assume the EoS parameters $\omega_{i}=p_{i}/\rho_{i}$, where $i=M,T$ for matter and torsion components, respectively. From (\ref{TFRW1}) and (\ref{TFRW2}) we write the continuity equation of the universe
$$\dot{\rho}_{c}+3H(\rho_{c}+p_{tot})=0.$$
It is convenient to introduce the effective EoS
$$\omega_{tot}=\sum \Omega_{i}\omega_{i},$$
where $\Omega_{i}=\rho_{i}/\rho_{c}$. We require the matter continuity equation
$$\dot{\rho}_{M}+3H(1+\omega_{M})\rho_{M}=0,$$
where matter EoS parameter $\omega_{M}=1/3$ for radiation and $\omega_{M}=0$ for dust. Consequently, we write the continuity equation of the torsion contribution as
$$\dot{\rho}_{T}+3H(1+\omega_{T})\rho_{T}=0,$$
where the torsion EoS parameter, using (\ref{Tor_dens}) and (\ref{Tor_press}), is given by
\begin{eqnarray}\label{Tor_EoS_par0}
\nonumber    \omega_{T}&=&p_{T}/\rho_{T}=-1+\frac{2}{3}\left[(1-f_{T}+12H^2f_{TT})\dot{H}\right.\\
\nonumber    &-&\left.{(1-f_{T}-12H^2 f_{TT})k/a^2}\right]/\\
&&{\left[f/6-(1+2f_{T})H^2-k/a^2\right]}.
\end{eqnarray}
\subsection{Reconstructing $f(T)$ from inflationary scene}\label{S4S1}
As shown in Section \ref{S2} that de Sitter related Hubble parameter directly to the cosmological constant by setting $\rho_{M}=p_{M}=0$ in Einstein's field equations. In this sense, the cosmological constant in de Sitter universe is sometimes thought of as the energy density of empty space. On the other hand, it is known that the choice of $f(T)=constant$ enforces the $f(T)$ field equations to produce a cosmological constant. In this work we, alternately, use the de Sitter approach by assuming a vanishing energy-momentum tensor $\mathcal{T}^{\mu \nu}$ $=$ $0$ of an empty space not by setting $f(T)=constant$. Thus we can reconstruct $f(T)$ in the absence of matter that can give rise to a de Sitter solution ($\rho_{M}=0$, $p_{M}=0$) by solving equations (\ref{dens1}) and (\ref{press1}). As $f(T)$ in FRW spacetime is a function of time $f(T \rightarrow t)$, one easily can show that
\begin{equation}\label{Fd2T}
  f_{T} = \dot{f}/\dot{T},~~  f_{TT} = \left(\dot{T} \ddot{f}-\ddot{T} \dot{f}\right)/\dot{T}^3.
\end{equation}
Substituting from (\ref{Fd2T}) into (\ref{dens1}) and (\ref{press1}), then by solving the system we get: The scale factor as
\begin{equation}\label{scale factor}
    a(t)=a_{0} e^{H_{0} (t-t_{0})},
\end{equation}
and the $f(T)$ as
\begin{equation}\label{fT}
    f(t)=\Lambda \exp{\left(\frac{-k e^{-2H_{0}(t-t_{0})}}{2a_{0}^{2}H_{0}^{2}}\right)},
\end{equation}
where $\Lambda$ and $a_{0}$ are constants of integration with an initial condition $H_{0}:=H(t_{0})$. In FRW cosmology the scale factor is the only dynamical variable, if we restrict ourselves to the GR. So we expect three different forms of the scale factor according to the choice of the sectional curvature $k$. In contrast, the $f(T)$ theories have two dynamical functions, that are the scale factor and the $f(T)$. In this model the scale factor (\ref{scale factor}) is independent of the value of $k$ and the universe in its vacuum state behaves initially same way in the three models $k=0 \pm 1$. However, different sectional curvatures affect the $f(T)$ evolution so that the theory in the flat case produces just a de Sitter cosmology. Although, the scale factor is independent of the value of $k$, the $f(T)$-theory can be thought of as an effective cosmological constant in the non-flat cases. Substituting from the vierbein (\ref{tetrad}) into (\ref{Tor_sc}), we found a non-vanishing value of the torsion scalar
\begin{equation}\label{Tsc}
T=-6H_{0}^2\left(1-\frac{k e^{-2 H_{0} (t-t_{0})}}{a_{0}^2 H_{0}^{2}}\right).
\end{equation}
The above expression enables to rewrite (\ref{fT}) as
\begin{equation}\label{fv}
f(T)=\Lambda e^{-\frac{T+6H_{0}^{2}}{12 H_{0}^{2}}}.
\end{equation}
In Section \ref{S2}, it has been shown that the empty space ($\mathcal{T}^{\mu \nu}=0$) requires vanishing of the Ricci scalar. This leads finally to accept the cosmological constant in the GR field equations to compensate the effect of the vacuum density. In $f(T)$ gravity, the torsion uniquely plays the main role in the teleparallel geometry so that the vanishing of the torsion tensor (\ref{q4}) implies the spacetime to be Minkowiskian. In our investigation here, unlike the case of the GR, fortunately the vanishing of the material distribution in the $f(T)$ does not imply a vanishing torsion scalar field (\ref{Tsc}). This enables us to investigate the vacuum energy density and explaining it by the torsion contribution. In flat space the teleparallel torsion scalar (\ref{Tsc}) reads $T=-6H_{0}^{2}$. Consequently, the equation (\ref{fv}) reads $f(T)=\Lambda$ which behaves as $\Lambda$ de Sitter universe. However, the non flat cases provide a dynamical evolution of the torsion contribution which serves as an effective cosmological constant. This leads us to consider the torsion as an intrinsic property of our spacetime itself!
\subsection{Missing ingredient density parameter}\label{S4S2}
Using the scale factor (\ref{scale factor}) the Hubble and the deceleration parameters become
\begin{equation}\label{Hubble}
    H=H_{0}=const.,~q=-1.
\end{equation}
Combining (\ref{scale factor}) and (\ref{Hubble}) gives a de Sitter universe. Actually, the de Sitter universe is invariant under time and space translations so that it does not allow the universe to evolve. So de Sitter universe could be useful at late accelerating expansion of the universe. But the early accelerating expansion cannot be exactly a de Sitter. The general relativistic treatment, which is characterized by the scale factor only, leads to a de Sitter universe. Fortunately, We are saved by introducing another dynamical function, $f(T)$, powered by the evolution of the curvature density parameter
\begin{equation}\label{curv_dens_para}
    \Omega_{k}= -\frac{k e^{-2H_{0}(t-t_{0})}}{a_{0}^{2} H_{0}^{2}}.
\end{equation}
In the GR treatment of FRW model, the curvature density parameter evolves much faster than other density parameters, which leads the universe to be more and more curved. On the other hand, all observations show that the present universe is almost flat. So it is believed that the best match between observations and GR is by taking the space to be initially flat, i.e. $k=0$. However, the observations provide a tiny value of $\Omega_{k}\sim O(10^{-5})$ but not zero \cite{1303.5082}. This could be due to a small initial scale factor or a large Hubble values. In this theory, the obtained $f(T)$ might give a backyard to relax this tough condition. Moreover, the cosmological constant problem might impose another tough condition by assuming some species with a positive vacuum energy are compensated by others with negative contributions up to $120$ decimal places \cite{V2006}. So the effective precession is not a trivial issue.

According to (\ref{FRW3}), the absence of the material distribution derives the curvature density parameter to be a unit. However, it vanishes at spatially flat spacetime as it should! Even in the non-flat cases, the curvature density parameter decays exponentially to zero. Although this solution contributes perfectly to solve the flatness problem, it shows that there is a missing ingredient to hold the total density parameter at unity
\begin{equation}\label{missing}
\Omega_{k}+\Omega_{(?)}=1,
\end{equation}
where $\Omega_{(?)}$ expresses the missing ingredient of the vacuum that we need to investigate. Most cosmologists assume the flat model seeking a perfect match with inflation requirements. However, taking the spatial flatness as a firm prediction of inflation is not quite accurate. It is easy to show that the smallness of $\Omega_{k}$ can be a result of $k\ll a^{2}H^{2}$ as expected at early universe. Moreover, it has been shown that even with a relatively large curvature parameter one can gain all advantages of inflation. Recent work has shown that the average of $|\Omega_{k}|\lesssim 0.15$ at $1\sigma$ confidence \cite{FMR15}. Also, it has been shown that a recognizable correlation between the modified growth parameters and $\Omega_{k}$ where $|\Omega_{k}|\geq 0.05$. All these evidences show that the spatial curvature must be included in the analysis with other cosmological parameters \cite{DI12,ZZCZ14}.\\

On the other hand, what does the missing ingredient in (\ref{missing}) tell? We need it to oppose the curvature density parameter. On another word, we need it to act in such away to compensate $\Omega_{k}$ holding the total of the density parameters at unity! Actually, this is also supported by seeing the density parameters as probabilities that must meet the unity when they cover the whole space. Fortunately, we are saved by the torsion contribution to the Friedmann equations!
\subsection{Torsion contribution}\label{S4S3}
Using the pairs $\left(a(t),~ f(T)\right)$ as given by (\ref{scale factor}) and (\ref{fT}) in (\ref{dens1}) and (\ref{press1}), it is easy to verify the empty space conditions as $\rho_{M}=0$ and $p_{M}=0$. However, using them in (\ref{Tor_dens}) and (\ref{Tor_press}), we evaluate the torsion gravity contribution to the density and pressure, respectively, as
\begin{equation}\label{Tor_den2}
  \rho_{T}=\frac{3H_{0}^{2}}{8\pi}\left(1+\frac{k}{a_{0}^2 H_{0}^{2}} e^{-2H_{0}(t-t_{0})}\right).
\end{equation}
\begin{equation}\label{Tor_press2}
  p_{T}=-\frac{3H_{0}^{2}}{8\pi}\left(1+\frac{k}{3a_{0}^2 H_{0}^2} e^{-2H_{0}(t-t_{0})}\right).
\end{equation}
The flat universe model shows that the effective torsion density and pressure produce the cosmological constant perfectly ($p_{T}=-\rho_{T}$). Now we can define the new parameter $\Omega_{T}:=\frac{\rho_{T}}{\rho_{c}}=\frac{\rho_{T}}{3H^2/8\pi}$ to represent the \textit{torsion density} parameter. Using equation (\ref{Tor_den2}) the torsion density parameter can be expressed as
\begin{equation}\label{Tor_dens_param}
    \Omega_{T}=1+\frac{k}{a_{0}^2 H_{0}^{2}}e^{-2 H_{0} (t-t_{0})}.
\end{equation}
It is clear that the above expression derives the total density parameter always to unity as $\Omega_{k}+\Omega_{T}=1$. So it clarifies the nature of the missing ingredient density parameter of (\ref{missing}) as the torsion density parameter.
\begin{figure}[t]
\centering
\subfigure[closed universe]{\label{fig1a}\includegraphics[scale=.2]{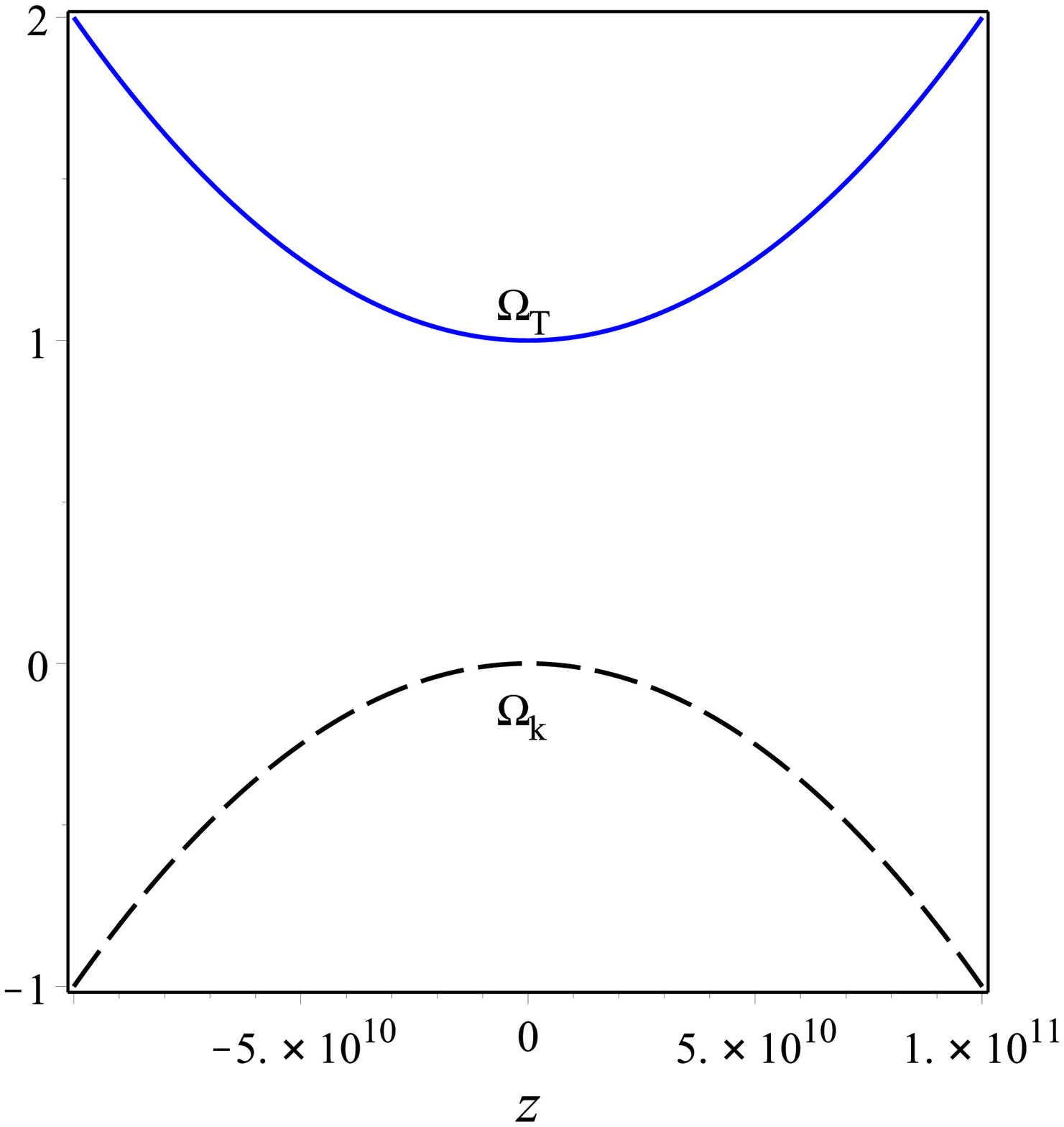}}
\subfigure[open universe]{\label{fig1b}\includegraphics[scale=.2]{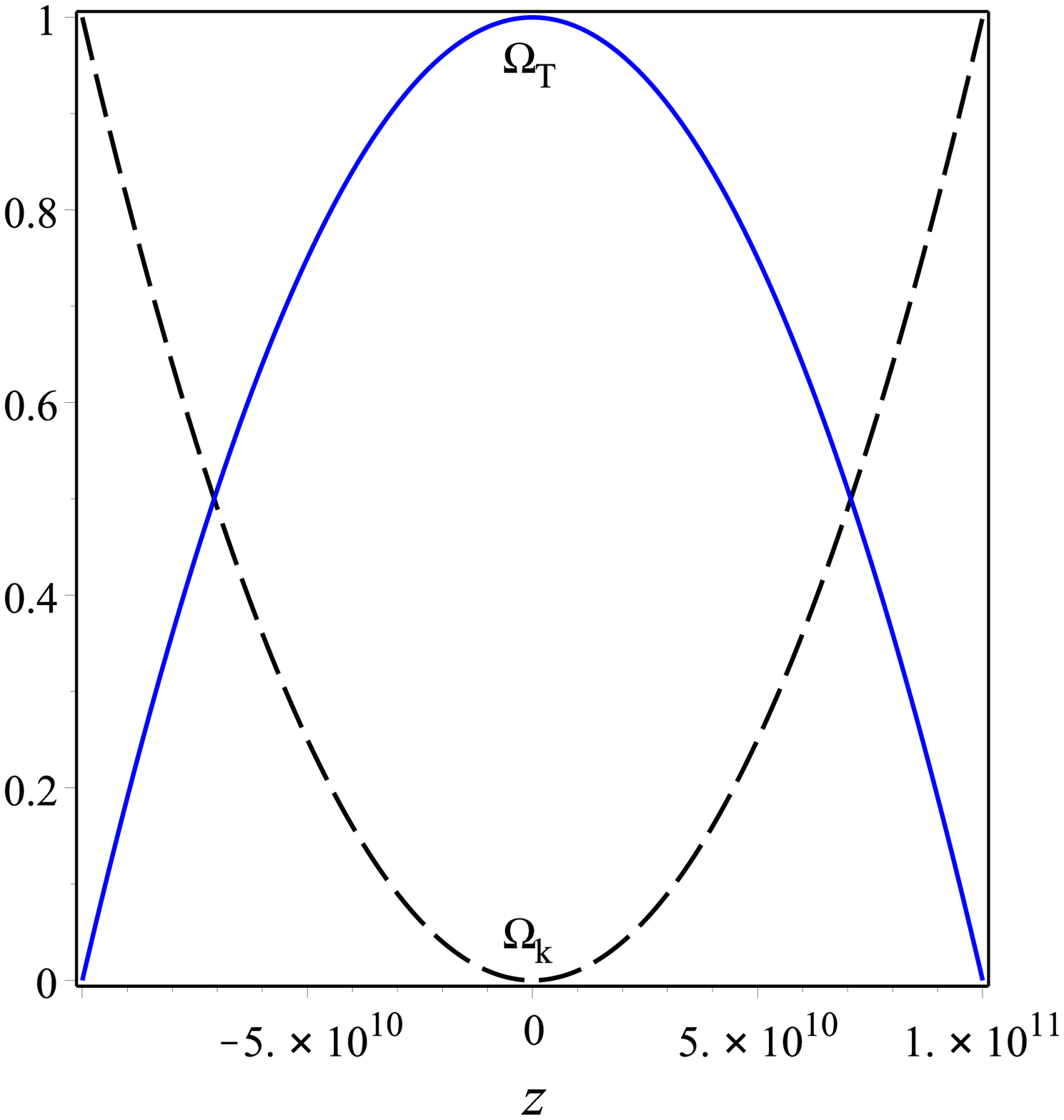}}
\caption{The evolution of the curvature and torsion density parameters vs the redshift $z$ as given by (\ref{curv_dens_para_Z}) and (\ref{Tor_dens_para_Z}):
\subref{fig1a} The closed universe model shows no overlap of the parameters;
\subref{fig1b} The open universe model shows a transition from curvature density domination to torsion density of period $-1-\frac{\sqrt{2}}{2}H_{0}\leqslant z \leqslant -1+\frac{\sqrt{2}}{2}H_{0}$ with an expected future transition at $z=-1-\frac{\sqrt{2}}{2}H_{0}$. The Hubble parameter has been chosen as $H_{0}=10^{10}$.}
\label{Fig1}
\end{figure}
We can redefine the cosmological parameters in terms of the redshift $z$. This can be achieved by using (\ref{scale factor}) into the redshift-scale factor relation $z=\frac{a_{*}}{a}-1$, where $a_{*}=1$ at present time. Consequently, we can express the time as a function of the redshift as
$$t=t_{0}+\frac{1}{H_{0}}\ln\left(\frac{1/a_{0}}{1+z}\right).$$
Then the curvature and torsion density parameters (\ref{curv_dens_para}) and (\ref{Tor_dens_param}) can be expressed, respectively, in terms of $z$ as given below
\begin{equation}\label{curv_dens_para_Z}
    \Omega_{k}(z)=-\frac{k}{H_{0}^{2}}(1+z)^2,
\end{equation}
\begin{equation}\label{Tor_dens_para_Z}
    \Omega_{T}(z)=1+\frac{k}{H_{0}^{2}}(1+z)^2.
\end{equation}
The evolution of the density parameters shows a battle between the vacuum Titans: the curvature and the torsion density parameters. Using (\ref{curv_dens_para_Z}) and (\ref{Tor_dens_para_Z}) enables to determine the redshift of the equality $\Omega_{k}=\Omega_{T}$ as $z=-1\pm \frac{H_{0}\sqrt{-2k}}{2k}$. The plot of the open universe model, where $k=-1$, in Figure \ref{Fig1} shows possible past transition period from curvature dominant universe to torsion domination epoch at $z=-1+\frac{\sqrt{2}}{2}H_{0}$ with expected another transition at $z=-1-\frac{\sqrt{2}}{2}H_{0}$. This period is independent of the choice of $a_{0}$ but depends only on the value of Hubble parameter $H_{0}$.
\section{Decaying $f(T)$-gravity in non-flat universe}\label{S5}
Many major challenges in physics today are present. What does power the accelerated expansion of our universe? Why do we accept using the cosmological constant to explain the vacuum energy? Even if we accept this, why does our universe so fine-tuned to a tiny positive value of the cosmological constant today? What mechanism enforces the large Planck vacuum density to its present tiny value? In this work, we argue possible explanation by reconstructing an $f(T)$ gravity theory in the absence of matter. We mentioned in the previous section that $f(T)$ of the flat space is just a cosmological constant. Otherwise, we have different scenarios. It has been shown that the observations suggest a non zero curvature density $\Omega_{k}\sim O(10^{-5})$, which implies a non flat sectional curvature $k \neq 0$. As shown by equation (\ref{curv_dens_para}) that the curvature density parameter begins with a large value at some early time $t_{i}$ after Planck time $t_{p}$, then it decays to a tiny value at a later time $t_{f}$. So it is more convenient to reexpress some functions in terms of the curvature density parameter. Using (\ref{curv_dens_para}) the torsion scalar (\ref{Tsc}) can be rewritten as
\begin{figure}[t]
\centering
\subfigure[closed universe]{\label{fig2a}\includegraphics[scale=.2]{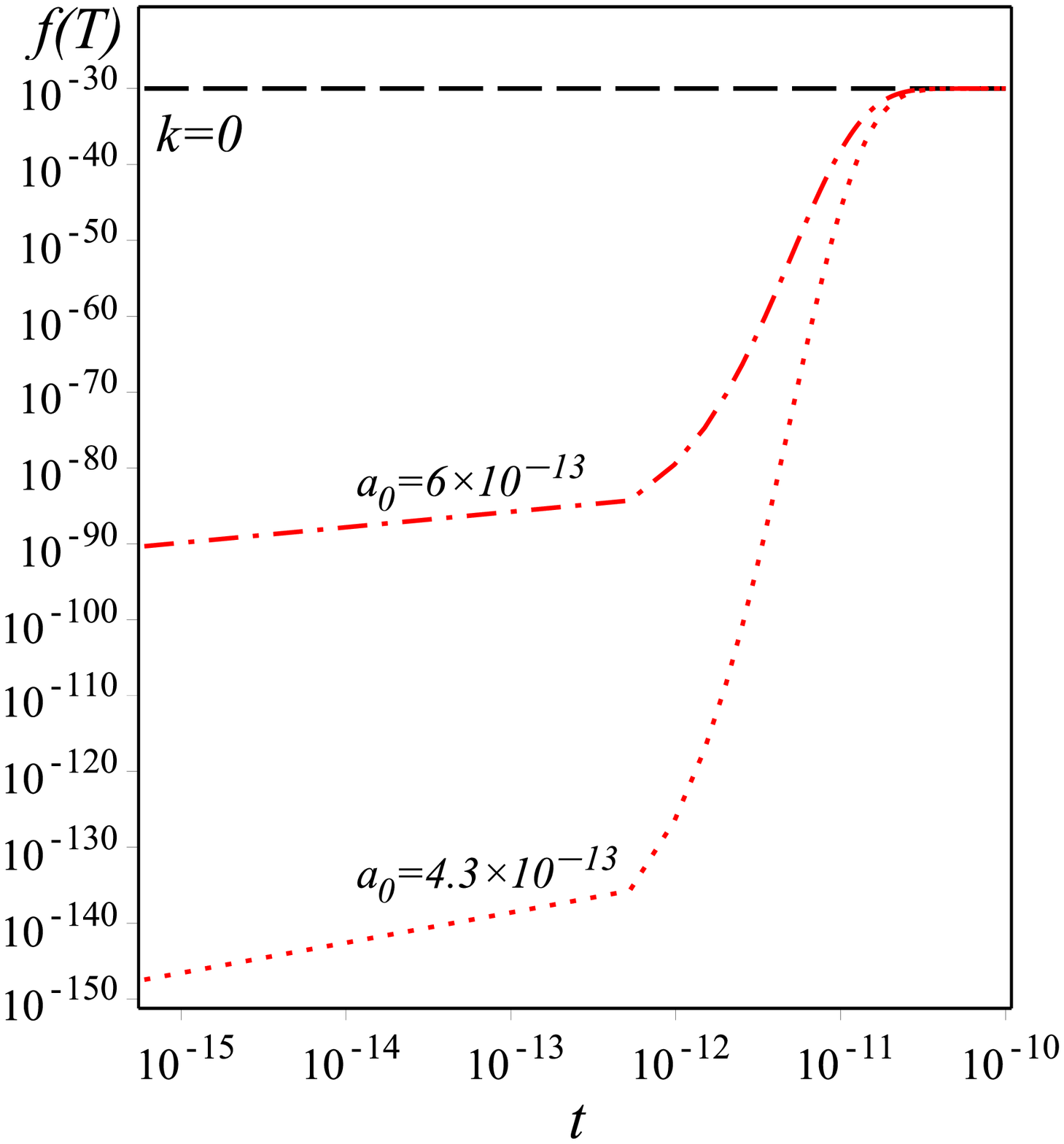}}\hspace{2pt}
\subfigure[open universe]{\label{fig2b}\includegraphics[scale=.2]{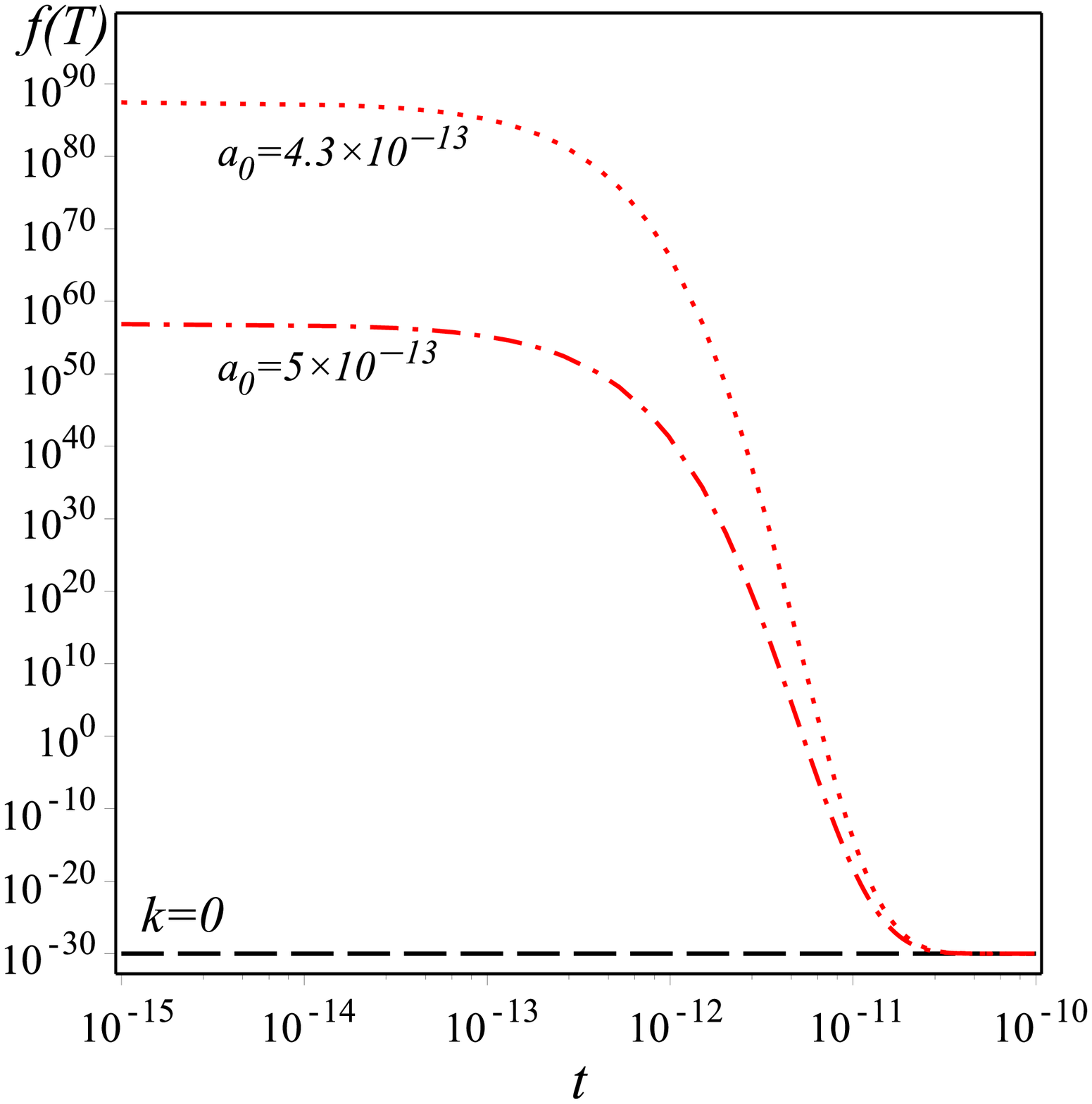}}
\caption{The evolution of the $f(T)$ gravity (\ref{fT}):
\subref{fig2a} The $f(T)$ grows exponentially in the closed universe model;
\subref{fig2b} The $f(T)$ decays exponentially in the open universe model. The initial conditions have been chosen as $t_{0}=t_{\textmd{p}}=10^{-44}~s$. Both show a sudden change at time $\sim 10^{-12}~s$. The open universe model shows a more realistic behaviour, whereas $f(T)$ decays from an initial large value of $\sim 2.76 \times 10^{87}s^{-2}$ to the small present value of the cosmological constant $\sim 10^{-30}~s^{-2}$.}
\label{Fig2}
\end{figure}
\begin{equation}\label{Tscalar}
   T=-6H_{0}^2(1+\Omega_{k}),
\end{equation}
similarly, using (\ref{fT}) the $f(T)$ can be rewritten as
\begin{equation}\label{fTk}
    f(T)=\Lambda e^{\Omega_{k}/2}.
\end{equation}
The above expressions relate the cosmic evolution to the decaying behaviour of the curvature density parameter $\Omega_{k}$. Accordingly, we describe qualitatively the evolution of (\ref{curv_dens_para}), (\ref{Tscalar}) and (\ref{fTk}) as
$$t \sim t_{f} \Rightarrow \left\{\begin{array}{l}
       \Omega_{k}\sim 0, \\
       T \sim -6H_{0}^{2}, \\
       f(T) \sim \Lambda.
\end{array}\right.$$
Now we investigate in particular the evolution of $f_{\textmd{\tiny vac.}}(T)$ quantitatively. Taking suitable values for the constants in (\ref{fT}) to match the early time of the empty universe. So we take an initial scale factor $a_{0}=4.3 \times 10^{-13}$, Hubble constant $H_{0}=10^{10}~s^{-1}$, initial time at Planck time $t_{0}=t_{p}=10^{-44}~s$. Also, as $f(T)\sim \Lambda$ at the limit $t\sim t_{f}$, we take the constant $\Lambda=10^{-30}~s^{-2}$ to match the measured value of the cosmological constant by the cosmological and astrophysical observations. The plots of Figure \ref{Fig2} show a sudden transition, in $f(T)$, has a short period from $t_{p}<t_{i}\lesssim 10^{-12}~s$ to $t_{f}\sim 10^{-10}~s$, such that
\begin{equation}\label{Adecay}
    \left[\frac{f(T)}{\Lambda}\right]_{t<t_{i}}\sim 10^{\pm 118},~ \textmd{while}
    \left[\frac{f(T)}{\Lambda}\right]_{t>t_{f}}\sim 1.
\end{equation}
We discuss separately each of the non flat models:

(\textbf{i}) In the closed universe model, Figure \ref{Fig2}\subref{fig2a} shows an exponential growth of $f(T)$ by about $118$ orders of magnitude. This requires a very small initial value of $f(T) \sim 3.63\times 10^{-148}~s^{-2}$ to match the present observed value of the cosmological constant.

(\textbf{ii}) The more interested case is the open universe model which contributes to solve the problem of the cosmological constant. Equation (\ref{fT}) indicates that $f(T)$ has a decaying behaviour, so that $f(T): \sim 2.76 \times 10^{87}~s^{-2} \mapsto 10^{-30}~s^{-2}$ as $t: t_{p} \mapsto t_{f}$. The plots of Figure \ref{Fig2}\subref{fig2b} show that the time interval of this decay is at $t_{f} \sim 10^{-12}-10^{-10}~s$; then it fixes its value to the present value of the cosmological constant. The evolution of $f(T)$ of the open universe model can perform a large value at the very early universe, while it decays to the present measured value of the cosmological constant. Moreover, the theory can predict the $\sim 117$ orders of magnitude between the two values. Also, the exponential scale factor (\ref{scale factor}) strongly provides an inflationary cosmic scenario. As is well known, the scalar field models are the most successful description of the inflationary universe. These models suggest that the early cosmic expansion is powered by a short lifetime $0$-spin particles (inflaton), which derive the universe to the required entropy. These models become strongly recommended than ever, in the light of the Higgs bosons detection at the LHC. Interestingly, the $f(T)$ lifetime as predicted by this theory is $\sim 10^{-12}~s$. This result matches the prior range of the scalar field lifetime.
\section{Physics of torsion}\label{S6}
The above mentioned illustration might open many questions. What is the nature of the torsion scalar field? How dose it contribute to the early or to the later phases of the universe? What is possible sources of the torsion scalar field? Does it propagate or not? What does torsion scalar field responsible of? However, we will try to investigate these questions, some of them are beyond our research here.
\subsection{Torsion equation of state}\label{S6S1}
\begin{figure}[t]
\centering
\subfigure[negative Hubble regime]{\label{fig3a}\includegraphics[scale=.2]{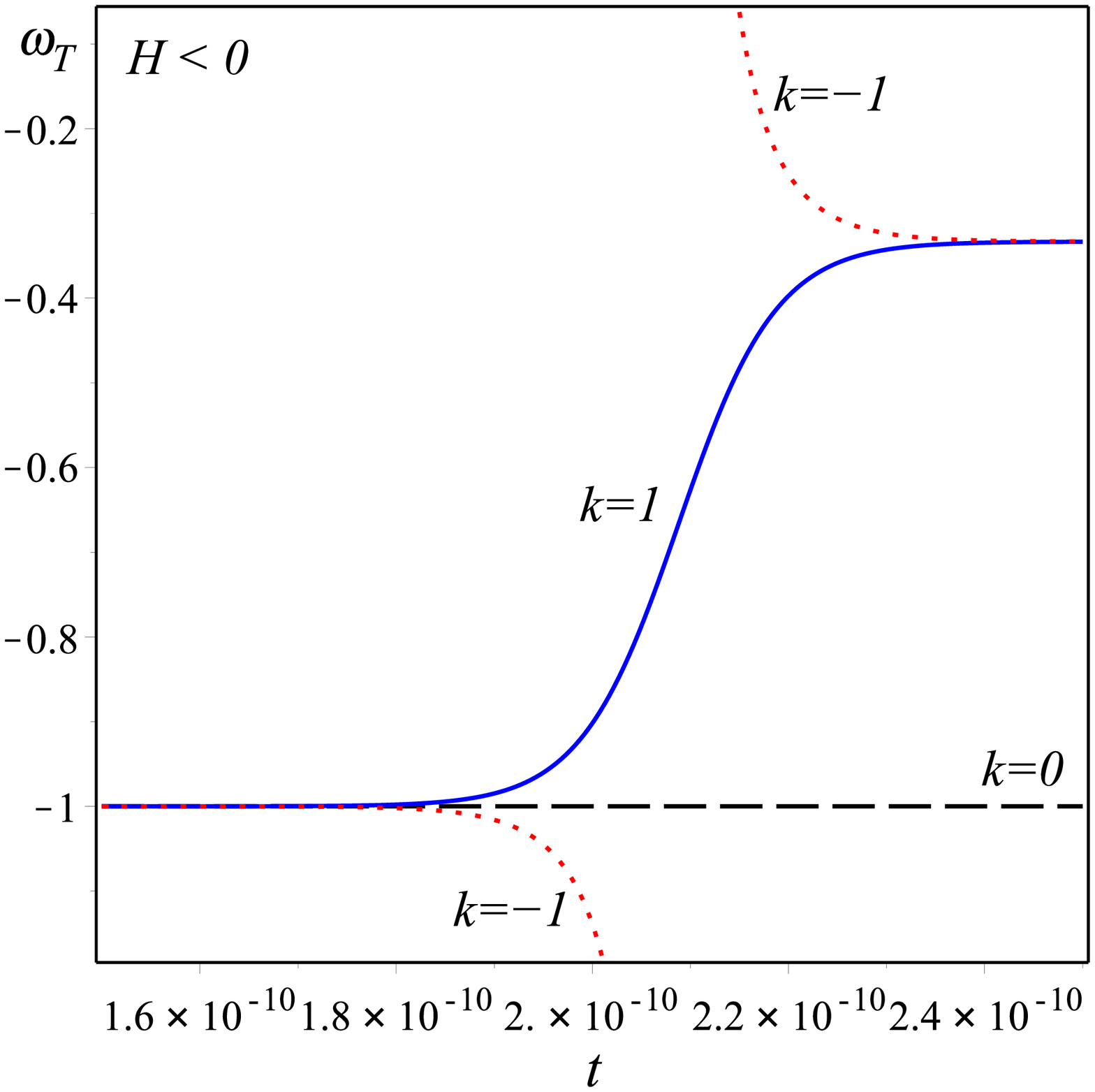}}\hspace{2pt}
\subfigure[positive Hubble regime]{\label{fig3b}\includegraphics[scale=.2]{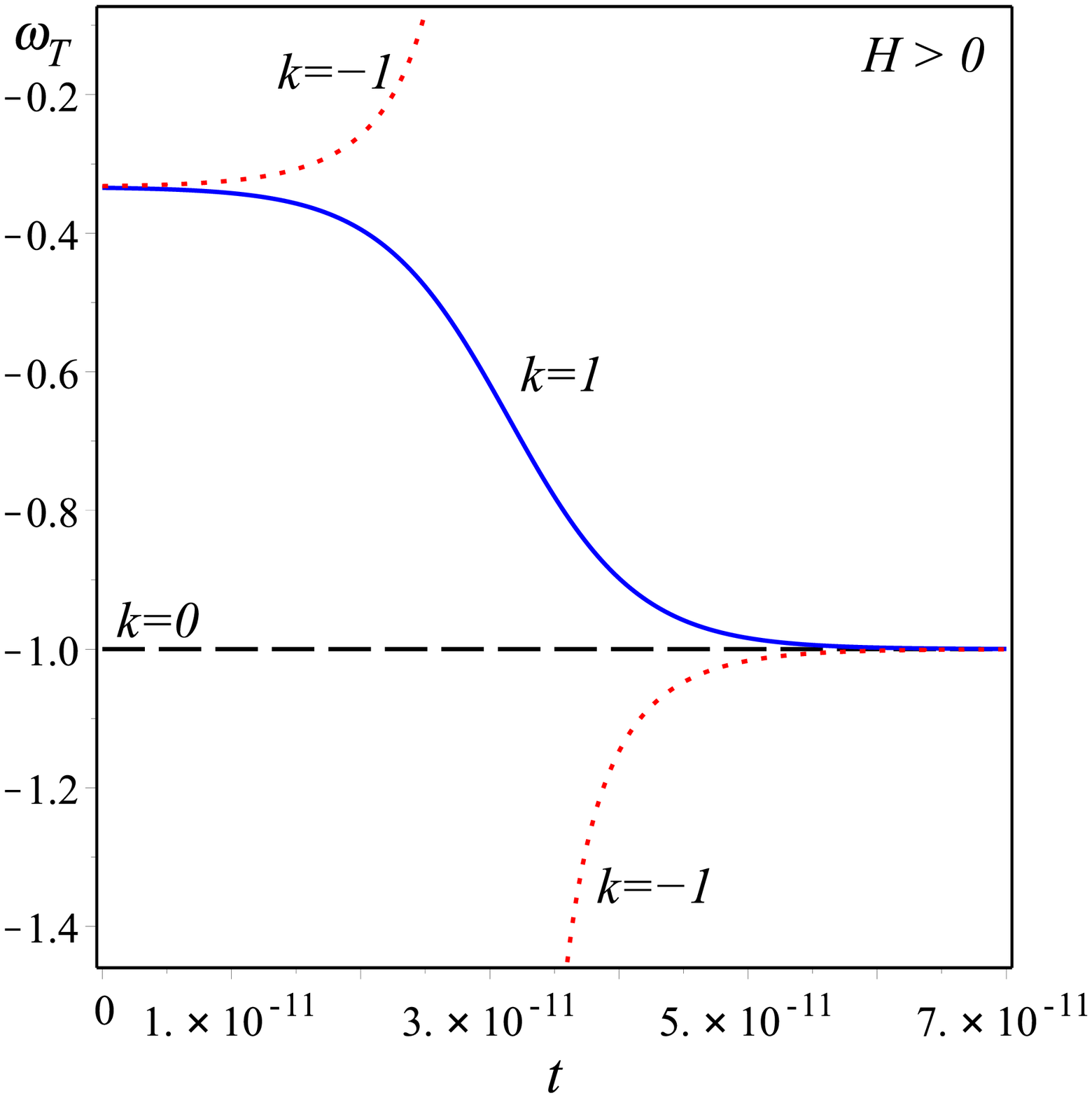}}
\caption{Evolution of the effective torsion fluid EoS (\ref{Tor_EoS_par}) vs. time:
\subref{fig3a} The torsion vacuum fluid EoS in the negative Hubble regime shows a tracker field behaviour which has a quintessence asymptotic behaviour with EoS parameter $\omega_{T} \rightarrow -1/3$;
\subref{fig3b} The torsion vacuum fluid EoS in the positive Hubble regime shows an initial quintessence matter ($\omega_{T}=-1/3$) evolves to cosmological constant in non-flat models. The EoS parameter shows a sudden singularity at the open universe model, while it has a non-dynamical behaviour in the flat universe model fixed to the cosmological constant ($\omega_{T}=-1$).}
\label{Fig3}
\end{figure}

We organize this section to answer some of the above mentioned questions. We evaluate the EoS parameter of the torsion scalar field to reveal its nature. This can be done by using (\ref{Tor_den2}) and (\ref{Tor_press2}), we get a time dependent EoS as
\begin{equation}\label{Tor_EoS_par}
    \omega_{T}=\frac{p_{T}}{\rho_{T}}=-\frac{1}{3}\frac{3a_{0}^2H_{0}^2+ke^{-2H_{0}(t-t_{0})}}{a_{0}^2H_{0}^2+ke^{-2H_{0}(t-t_{0})}}.
\end{equation}
In non-flat models, the EoS evolves from $\omega_{T}: -1 \mapsto -\frac{1}{3}$ as time runs $t: t_{p} \mapsto t_{f}$ for negative Hubble spacetime, while the more physical case where Hubble parameter is positive the EoS evolves as $\omega_{T}: -\frac{1}{3} \mapsto -1$ as time runs as $t: t_{p} \mapsto t_{f}$. For $H>0$ spacetime, the $k=-1$ EoS shows a sudden singularity, however, it has a unified asymptotic behaviour matching the flat limit $\omega_{T}|_{k=0}=-1$. We summarize the EoS evolution as follows:
$$\omega_{T}=\left\{\begin{array}{lcl}
       k=0,~a_{0}, H_{0} \neq 0 &\Rightarrow & \omega_{T}=-1 \\
       k \neq 0,~H_{0} \leq 0 &\Rightarrow& \omega_T: -1 \mapsto -\frac{1}{3} \\
       k \neq 0,~H_{0} > 0 &\Rightarrow& \omega_T: -\frac{1}{3} \mapsto -1
\end{array}\right.$$
\begin{table}[b]
\caption{Torsion vacuum EoS (\ref{EoS_curv}) corresponds to the evolution of $\Omega_{k}$}
\label{T1}
\begin{tabular*}{\columnwidth}{@{\extracolsep{\fill}}cccc@{}}
\hline
EoS & $\Omega_{k}$ & universe& phase  \\
\hline
\multicolumn{1}{c}{$\omega_{T}<0$} & $\Omega_{k}<1$ & closed or open & inflation  \\
                                   & $\Omega_{k}>3$ & open & inflation \\
$\omega_{T}=0$ & $\Omega_{k}=3$ & open & dust \\
$\omega_{T}>0$ & $1<\Omega_{k}<3$ & open & radiation\\
$\omega_{T}=\infty$ & $\Omega_{k}=1$ & open & singularity\\
\hline
\end{tabular*}
\end{table}
Also, we need to mention the only case that the evolution of the torsion scalar field requires a sudden singularity is the case of the open universe model ($k=-1$). Different evolution scenarios are given in Figure \ref{Fig3}.
Also, its useful to express the torsion EoS in terms of $\Omega_{k}$. Substituting from (\ref{curv_dens_para}) into (\ref{Tor_EoS_par}); then
\begin{equation}\label{EoS_curv}
    \omega_{T}=-\frac{1}{3}\left(\frac{3-\Omega_{k}}{1-\Omega_{k}}\right).
\end{equation}
The above equation puts a constraint on the curvature density parameter $\Omega_{k} \neq 1$. Also, it shows the sensitivity of the torsion EoS to the evolution of the curvature density parameter. When $\Omega_{k}$ is large enough, i.e. the pre-inflation period, the torsion EoS $\omega_{T}\rightarrow -1/3$. The exponential scale factor (\ref{scale factor}) derives the universe to the opposite limit $\Omega_{k} \rightarrow 0$, i.e. spatially flat like, the torsion EoS $\omega_{T} \rightarrow -1$ by the end of inflation. So the universe approaches the de Sitter fate. This showing that the EoS of the torsion vacuum fluid is trapped in the interval $-1<\omega_{T}<-1/3$, which perfectly matches the requirements of the cosmic accelerating expansion. We summarize the results of (\ref{EoS_curv}) in Table \ref{T1}.
\subsection{Torsion potential of a scalar field}\label{S6S2}
We consider here the physical approach to form the torsion from a scalar field $\varphi(x)$. We follow the approach that has been purposed by \cite{XSH96}, by introducing sixteen fields $t^{\mu}{_{i}}$ that are called ``\textit{torsion potential}". These fields form a quadruplet basis vectors, so we write the following linear transformation:
\begin{equation}
\nonumber    h_{i}=t^{\mu}{_{i}}\partial_{\mu},~~h^{i}=t^{i}{_{\mu}}dx^{\mu},
\end{equation}
the torsion potential $t^{\mu}{_{i}}$ and its inverse satisfy
\begin{equation}
\nonumber    t=\textmd{det}(t^{\mu}{_{i}})\neq 0,~~t^{\mu}{_{i}}t^{i}{_{\nu}}=\delta^{\mu}_{\nu},~~t^{\mu}{_{i}}t^{j}{_{\mu}}=\delta^{j}_{i}.
\end{equation}
Then the torsion can be written as \cite{XSH96,HN15}
\begin{equation}\label{torsion_pot}
    T^{\alpha}{_{\mu \nu}}=t^{\alpha}{_{i}}\left(\partial_{\nu}t^{i}{_{\mu}}-\partial_{\mu}t^{i}{_{\nu}}\right).
\end{equation}
Generally, the torsion potential $t^{\mu}{_{i}}$ can be reformed by a physical scalar, vector or tensor fields. In the inflationary models the cosmic inflation is powered by a spin-$0$ matter, so it is more convenient to assume the case when the torsion potential is constructed by a scalar field $\varphi(x)$, sometimes it is called tlaplon field \cite{HRR78,Mukku:1979}, which serves as torsion potential. Thus we can reformulate the $f(T)$ theory in terms of tlaplon field with an FRW background, so we take
\begin{equation}
\nonumber    t^{i}{_{\mu}}=\delta^{i}_{\mu}e^{\sqrt{3/2}\varphi},~~t^{\mu}{_{i}}=\delta^{\mu}_{i}e^{-\sqrt{3/2}\varphi},
\end{equation}
where $\varphi$ is a non-vanishing scalar field. Then the torsion is expressed as
\begin{eqnarray}
    T^{\alpha}{_{\mu\nu}}&=&\sqrt{3/2}\left(\delta^{\alpha}_{\nu}\partial_\mu \varphi-\delta^{\alpha}_{\mu}\partial_\nu \varphi\right),\label{semi-symm-torsion}\\
    K^{\mu\nu}{_{\alpha}}&=&\sqrt{3/2}\left(\delta^{\nu}_{\alpha}\partial^\mu \varphi-\delta^{\mu}_{\alpha}\partial^\mu \varphi\right),\label{semi-symm-contortion}
\end{eqnarray}
where $\partial^\mu \varphi=g^{\mu \alpha} \partial_\alpha \varphi$. The above equations show that the torsion trace is proportional to the gradient of the tlaplon field $T_{\alpha}\propto \partial_{\alpha}\varphi$ which allows the torsion to propagate in vacuum. Also, this form of the torsion tensor has been used successfully to couple torsion minimally to any gauge field \cite{R74,HRR78,H90,HO01}. Using the above equations and (\ref{q5}), the teleparallel torsion scalar (\ref{Tor_sc}) can be written in terms of the scalar field $\varphi$ as
\begin{equation}\label{transf}
    T=-9\partial_\mu\varphi~ \partial^\mu\varphi.
\end{equation}
Using equation (\ref{transf}) we apply the procedure appeared in \cite{HN15} to map the torsion contribution in the Friedmann equations into the scalar field ($\rho_{T}\rightarrow \rho_{\varphi}$, $p_{T}\rightarrow p_{\varphi}$). This leads to reformulate the Friedmann equations of the torsion contribution as an inflationary background in terms of the scalar field $\varphi$. At the inflationary epoch the matter contribution can be negligible, so we consider the Lagrangian density of a homogeneous (real) scalar field $\varphi$ in potential $V(\varphi)$
\begin{equation}\label{lag_dens}
    \mathcal{L}_{\varphi}=\frac{1}{2}\partial_\mu \varphi~ \partial^\mu \varphi-V(\varphi).
\end{equation}
where the first term in the above equation represents the kinetic term of the scalar field, as usual, while $V(\varphi)$ represents the potential of the scalar field. The variation of the action with respect to the metric $g_{\mu \nu}$ enables to define the energy momentum tensor as
\begin{equation}
 \nonumber   \mathcal{T}^{\mu \nu} = \frac{1}{2}\partial^\mu \varphi~ \partial^\nu \varphi-g^{\mu \nu} \mathcal{L}_{\varphi}.
\end{equation}
The variation with respect to the scalar field reads the scalar field density and pressure respectively as
\begin{equation}\label{press_phi1}
    \rho_{\varphi}=\frac{1}{2}\dot{\varphi}^2+V(\varphi),~~ p_{\varphi}=\frac{1}{2}\dot{\varphi}^2-V(\varphi).
\end{equation}
The Friedmann equation (\ref{TFRW1}) of the non-flat models in absence of matter becomes
\begin{equation}\label{Hubble_sc}
H^2=\frac{8\pi}{3}\left(\frac{1}{2}\dot{\varphi}^2+V(\varphi)\right)-\frac{k}{a^{2}}.
\end{equation}
Using (\ref{press_phi1}) it is an easy task to show that continuity equation of the torsion contribution
\begin{equation}
\nonumber \dot{\rho}_{T\rightarrow \varphi}+3H(\rho_{T\rightarrow \varphi}+p_{T\rightarrow \varphi})=0,
\end{equation}
can be mapped to the Klein-Gordon equation of homogeneous scalar field in the expanding FRW universe
\begin{equation}
\nonumber    \ddot{\varphi}+3H\dot{\varphi}+V'(\varphi)=0,
\end{equation}
where the prime denotes the derivative with respect to the scalar field $\varphi$. In conclusion, equation (\ref{transf}) enables to define a scalar field sensitive to the vierbein field, i.e. the spacetime symmetry. In addition, equation (\ref{press_phi1}) enables to evaluate an effective potential from the adopted $f(T)$ gravity theory. The mapping from the torsion contribution to scalar field fulfills the Friedmann and Klein-Gordon equations. So the treatment meets the requirements to reformulate the torsion contribution in terms of a scalar field without attempting conformal transformations.
\section{Torsion gravity and inflation}\label{S7}
As is well known, inflationary models relate the inflation beginning and ending to potential and kinetic energies domination, respectively, of one or more scalar fields. In the $f(T)$ vacuum theory, we study the compensation of the potential and the kinetic energies according to the evolution of the curvature density parameter. Using (\ref{Tscalar}) and (\ref{transf}) we write
\begin{equation}\label{kin1}
    \dot{\varphi}=\sqrt{\frac{2}{3}}H_{0}\sqrt{1+\Omega_{k}}.
\end{equation}
Substituting from (\ref{kin1}) into (\ref{Hubble_sc}); then the potential can be written as
\begin{equation}\label{V_curv}
    V(\Omega_{k})=\frac{1}{24\pi}H_{0}^{2}\left[9(1-\Omega_{k})-8\pi(1+\Omega_{k})\right].
\end{equation}
Also, the curvature density parameter (\ref{curv_dens_para}) rate of change can be given as
\begin{equation}\label{Omega_rate}
    \dot{\Omega}_{k}=-\frac{\Omega_{k}}{2 H_{0}}.
\end{equation}
Using (\ref{kin1}) and (\ref{Omega_rate}) the scalar field rate of change with respect to the curvature density is
\begin{equation}\label{phid_curv}
    \frac{d\varphi}{d\Omega_{k}}=\frac{\dot{\varphi}}{~\dot{\Omega}_{k}}
    =2\sqrt{\frac{2}{3}}H_{0}^{2}\frac{\sqrt{1+\Omega_{k}}}{\Omega_{k}}.
\end{equation}
\begin{figure}[t]
\centering
\includegraphics[scale=.4]{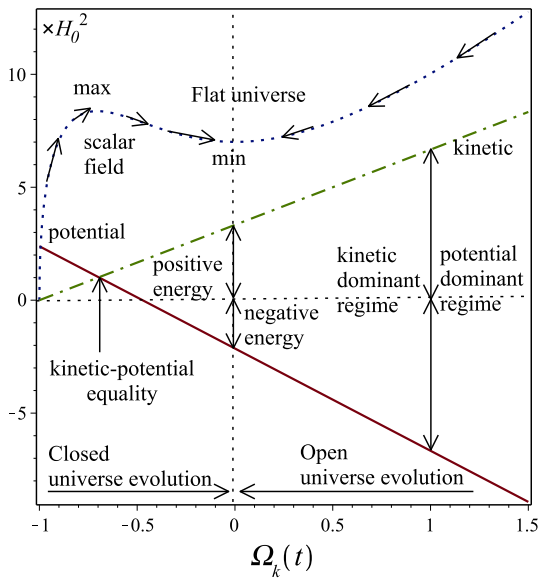}
\caption{Phase diagram of the kinetic, potential energies and the scalar field vs the curvature density parameter as given in (\ref{kin1}), (\ref{V_curv}) and (\ref{phi_curv}), respectively. The closed universe $-1<\Omega_{k}<0$. The flat universe $\Omega_{k}=0$. The open universe $\Omega_{k}>0$. The dot, dashdot and solid lines represent the scalar field, its kinetic and potential energies, respectively. The initial value has been taken as $\varphi_{0}=0$, the energy scale on the vertical axis is normalized to the $H_{0}^{2}$ scale.}\label{Fig4}
\end{figure}
Integrating the above equation, we have
\begin{equation}\label{phi_curv}
    \varphi(\Omega_{k})=\varphi_{0}+4\sqrt{\frac{2}{3}}H_{0}^{2}\left[\sqrt{1+\Omega_{k}}-\arctan{\sqrt{1+\Omega_{k}}} \right],
\end{equation}
where $\varphi_{0}$ is a constant of integration with a boundary condition $\Omega_{k}(t_{f})\sim 0$. Using (\ref{kin1}), (\ref{V_curv}) and (\ref{phi_curv}) the evolution of the scalar field and its kinetic and potential energies with respect to the curvature density parameter are represented in the plots of Figure \ref{Fig4}. In this $f(T)$ theory, the dynamical evolution is recognized only in non-flat regimes, we discuss separately the closed and the open universe models in the light of the plots of Figure \ref{Fig4}.
\subsection{The closed universe}\label{S7S1}
In this model the curvature density parameter $\Omega_{k}<0$, the scalar field (\ref{phi_curv}) restricts the curvature density parameter to $[-1,0]$ range. On the other hand, the closed universe model with $\Omega_{k}<-1$ allows only phantom (ghost) dark energy, i.e. negative kinetic energy. We reject this portion on the $\Omega_{k}$ timeline as is $-1/3<\omega_{T}<-1$ in the closed universe model.\\

Recalling Figure \ref{Fig1}\subref{fig1a}, the curvature density parameter evolution towards the flat limit $\Omega_{k}(t_{f})\sim 0$ is associated with a scalar field growth, Figure \ref{Fig4}, with a local maximum at $\Omega_{k}\sim -0.704$. After that it decays as evolving towards the flat universe limit. Also, at $\Omega_{k}\sim -1$ the scalar field potential is much greater than its kinetic energy, $V\gg \frac{1}{2}\dot{\varphi}^2$, which is suitable for cosmic inflation.\\

At $\Omega_{k}\sim -0.696$, the potential and kinetic energies reach their equality $V=\frac{1}{2}\dot{\varphi}^{2}$ allowing cosmic inflation to end. Remarkably, the kinetic-potential equality is just after the scalar field approaches its local maximum on the $\Omega_{k}$-timeline; then the results are consistent so far. During the period $\Omega_{k}=[-0.696,0]$ the scalar field decays towards its minimum at the flat universe limit as $\Omega_{k}(t_{f})\sim 0$ with kinetic dominant universe. From (\ref{Hubble_sc}) and (\ref{kin1}), obviously the potential is negative as $\Omega_{k}>\frac{9-8\pi}{9+8\pi} \sim-0.473$, the universe may enter a recollapse phase. At the flat universe limit the positive and negative energies almost compensate each other.
\subsection{The open universe}\label{S7S2}
In this model the curvature density parameter $\Omega_{k}>0$, the scalar field (\ref{phi_curv}) does not restrict the curvature density parameter to a particular range. On the other hand, the open universe model does not allow ghosts as the scalar field kinetic energy is always positive.\\

Recalling Figure \ref{Fig1}\subref{fig1b}, the curvature density parameter evolution towards the flat limit $\Omega_{k}(t_{f})\sim 0$ is associated with a scalar field decay, Figure \ref{Fig4}, towards the flat universe limit. Also, at $\Omega_{k}\gg 1$ the scalar field potential is much greater than its kinetic energy, $V\gg \frac{1}{2}\dot{\varphi}^2$, which is suitable for cosmic inflation.\\

At $\Omega_{k}= 1$, the potential and kinetic energies reach their equality $V=\frac{1}{2}\dot{\varphi}^{2}$ allowing cosmic inflation to end. The end of inflation is associated with a reheating process to transform the energy source of inflation into heat allowing transition to radiation dominated era with enough entropy. This result is in agreement with the results in Table \ref{T1}, where the positive EoS is allowed only at open universe as $\Omega_{k}\rightarrow 1$. Back to Figure \ref{Fig4}, the $0<\Omega_{k}< 1$ range allows a kinetic dominant universe which is suitable for galaxy formation. However, the universe will not enter a recollapse phase. Finally, the open universe model always seeking for a flat like limit allowing the positive and negative energies to compensate each other which is very similar to the end of inflation epoch with almost fixed scalar field background.
\subsection{Gravitational quintessence model}\label{S7S3}
In order to express the potential in terms of the scalar field, we use (\ref{V_curv}) and (\ref{phi_curv}); then the potential normalized to $H_{0}^{2}$ is
\begin{equation}\label{pot_phi}
    V(\varphi)=V_{0}+\frac{1}{2}\left(\varphi-\sqrt{2/3} \zeta(\varphi)\right)^{2},
\end{equation}
where $V_{0}=\frac{3}{8\pi}$, and $\zeta(\varphi)$ is at which $\tan^{2}{\zeta}=\frac{3}{2}\varphi^{2}+6\varphi \zeta + 6\zeta^{2}$. We take $\zeta$ small enough so that $\tan{\zeta}\approx \zeta+\frac{1}{3}\zeta^{3}+O(\zeta^{5})$; then the potential (\ref{pot_phi}) might read six different versions of the vacuum potential:
\begin{figure}[t]
\centering
\includegraphics[scale=.3]{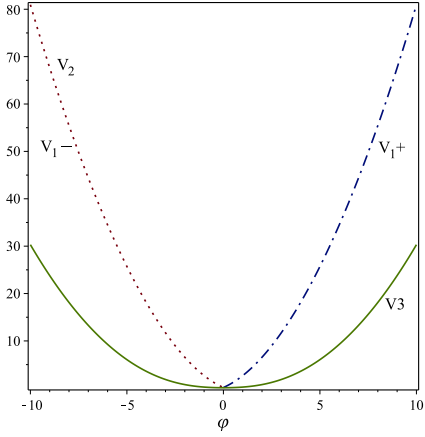}
\caption{The allowed real scalar field: The potentials $V_{1}$, $V_{2^{+}}$ and $V_{3}$ correspond to (\ref{V1}), (\ref{V2}) and (\ref{V3}), respectively.}\label{Fig5}
\end{figure}
\begin{equation}\label{V1}
    V_{1^{\pm}}=V_{0}+\frac{\varphi^{2}}{2}+\frac{1}{2}\sqrt[3]{\pm 4\varphi^{2}}+\sqrt[3]{\pm 2\varphi^{4}},
\end{equation}
where the $\pm$ denotes two different potentials. In the real scalar field case, the $V_{1^{+}}$ potential allows inflation only $\varphi>0$ plateau, while the potential $V_{1^{-}}$ allows inflation only at $\varphi<0$, see Figures \ref{Fig5}. If we consider the complex scalar field case, the projection of the potential of Figure \ref{Fig6}\subref{fig6a} in the real domain of the scalar field shows slow roll pattern, while potential projection in the imaginary scalar field domain appears as quadratic models. Also, we evaluate the potential
\begin{eqnarray}\label{V2}
\nonumber V_{2}&=&V_{0}+\frac{\varphi^{2}}{2}-\frac{\sqrt[3]{4}}{4}(\sqrt{3}\Im+1)\sqrt[3]{\varphi^{2}}\\
&&+\frac{\sqrt[3]{2}}{2}(\sqrt{3}\Im-1)\sqrt[3]{\varphi^{4}},
\end{eqnarray}
where $\Im=\sqrt{-1}$. The $V_{2}$ potential similar to the $V_{1^{-}}$ allows inflation only at $\varphi<0$ for the real scalar field case. The complex scalar field case are given in Figure \ref{Fig6}\subref{fig6b}. The projection in the real domain of the scalar field shows slow roll model, while its projection in the scalar field imaginary domain shows the old inflation model where potential shows a tunneling event from the high energy false vacuum to the true vacuum. Moreover, we evaluate
\begin{figure}[b]
\centering
\subfigure[The potential $V_{1^{\pm}}$]{\label{fig6a}\includegraphics[scale=.3]{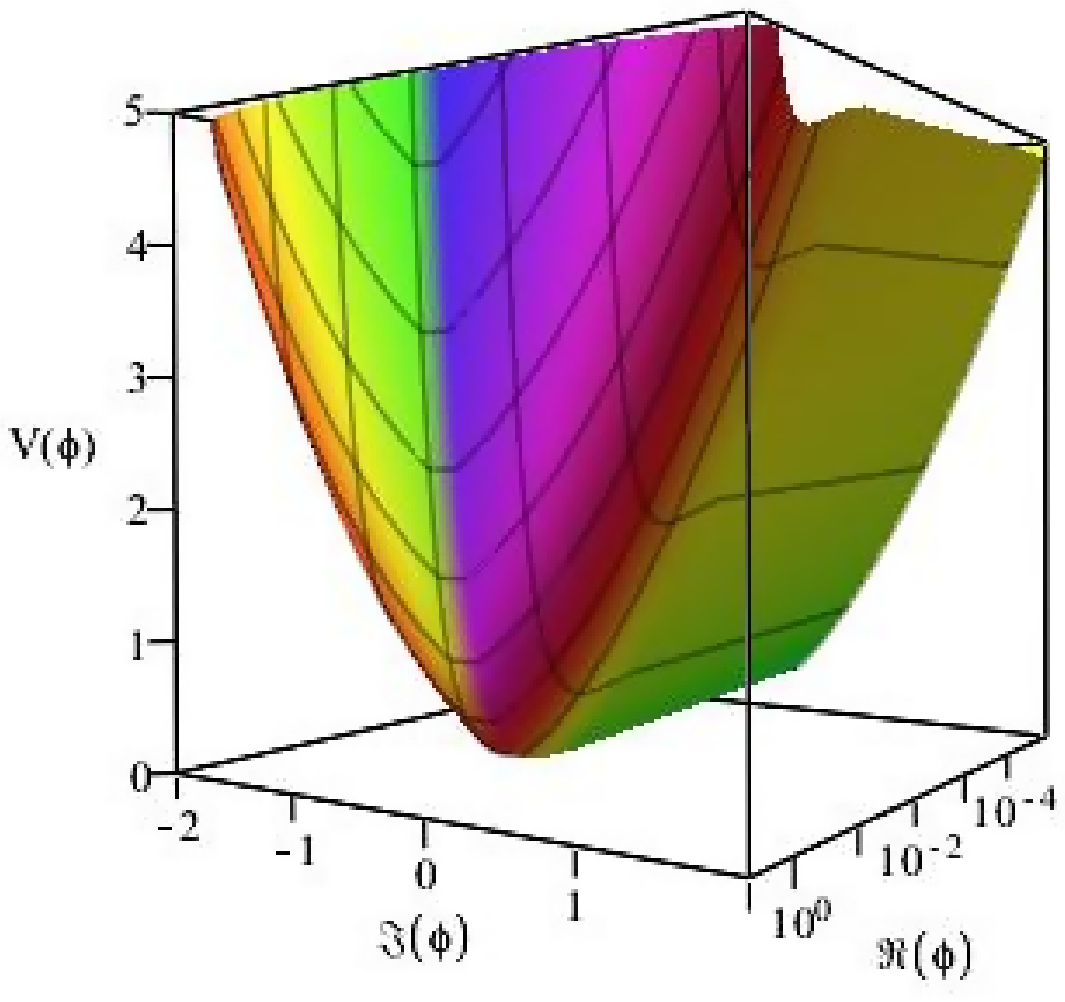}}
\subfigure[The potential $V_{2}$]{\label{fig6b}\includegraphics[scale=.3]{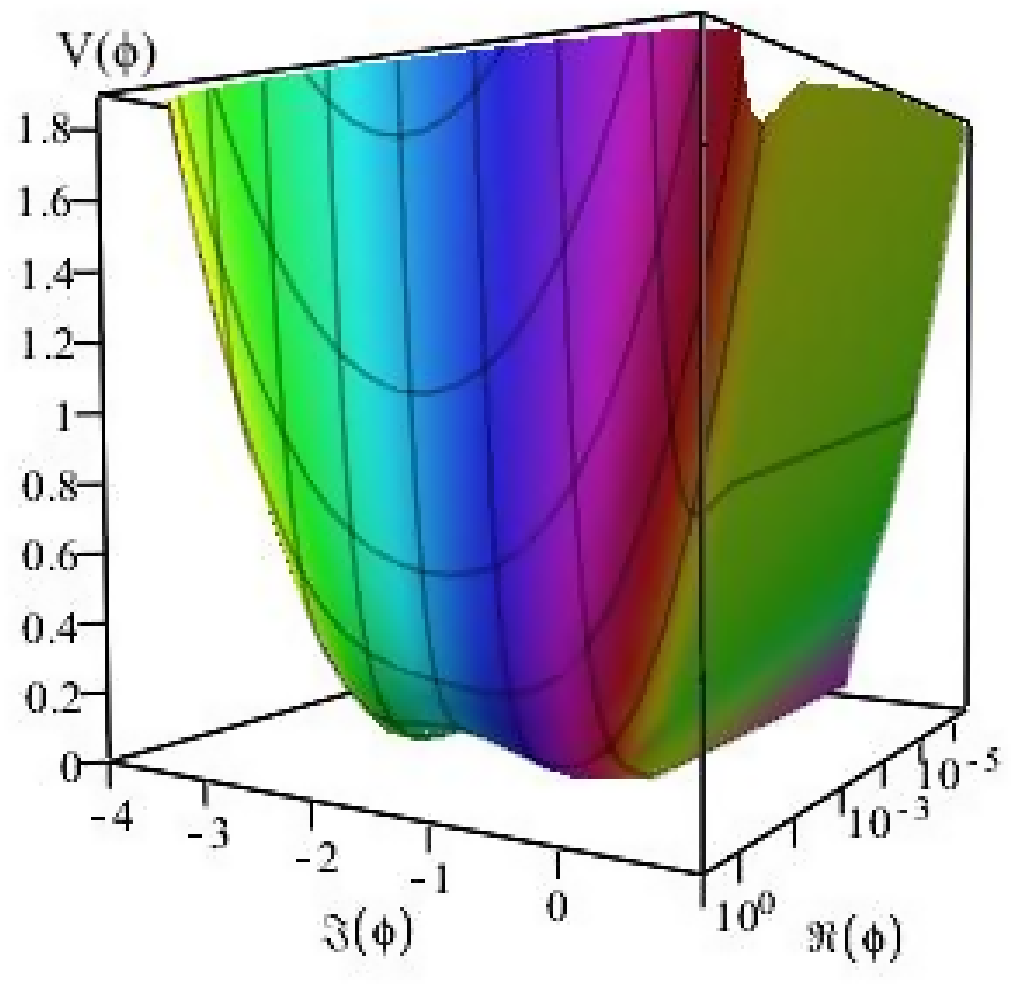}}\\
\subfigure[The potential $V_{3}$]{\label{fig6c}\includegraphics[scale=.3]{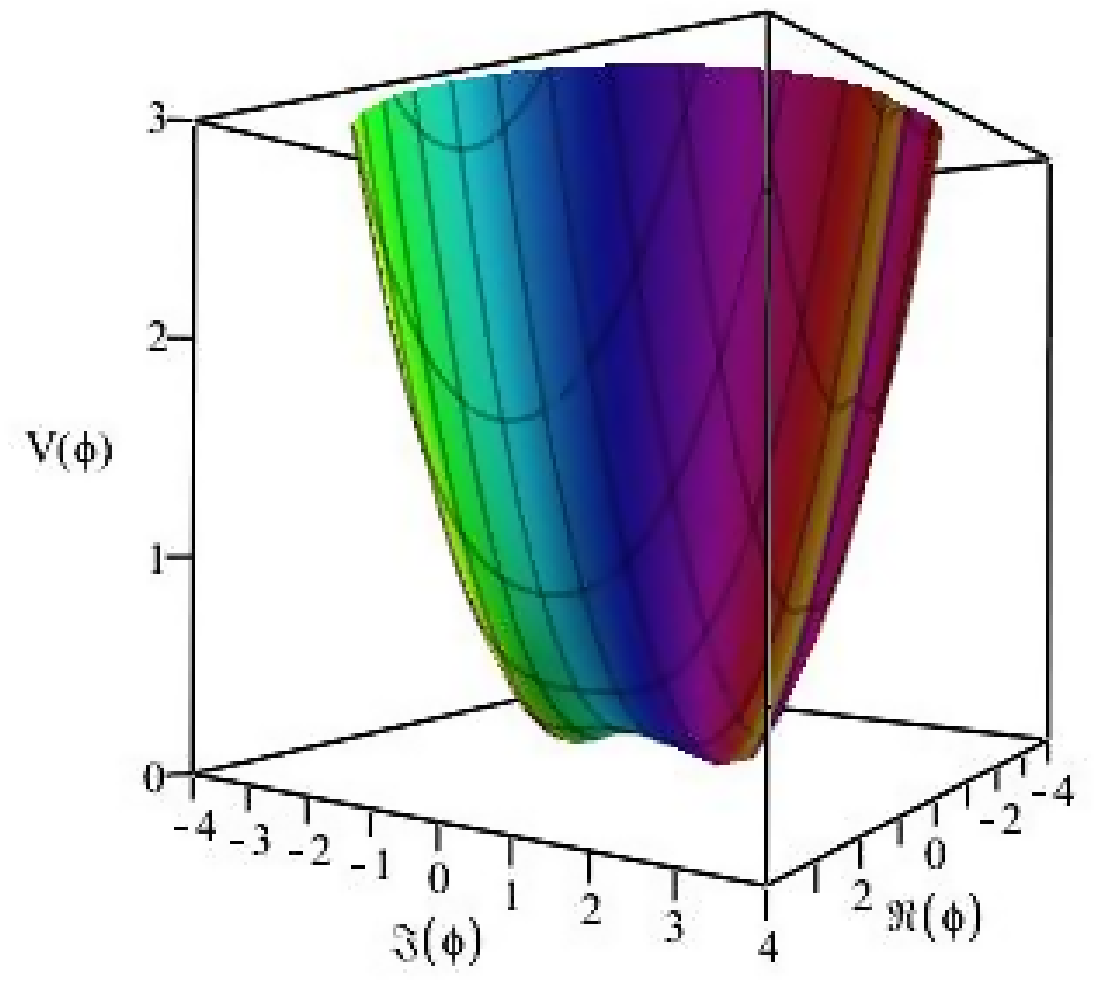}}
\subfigure[The potential $V_{4^{\pm}}$]{\label{fig6d}\includegraphics[scale=.3]{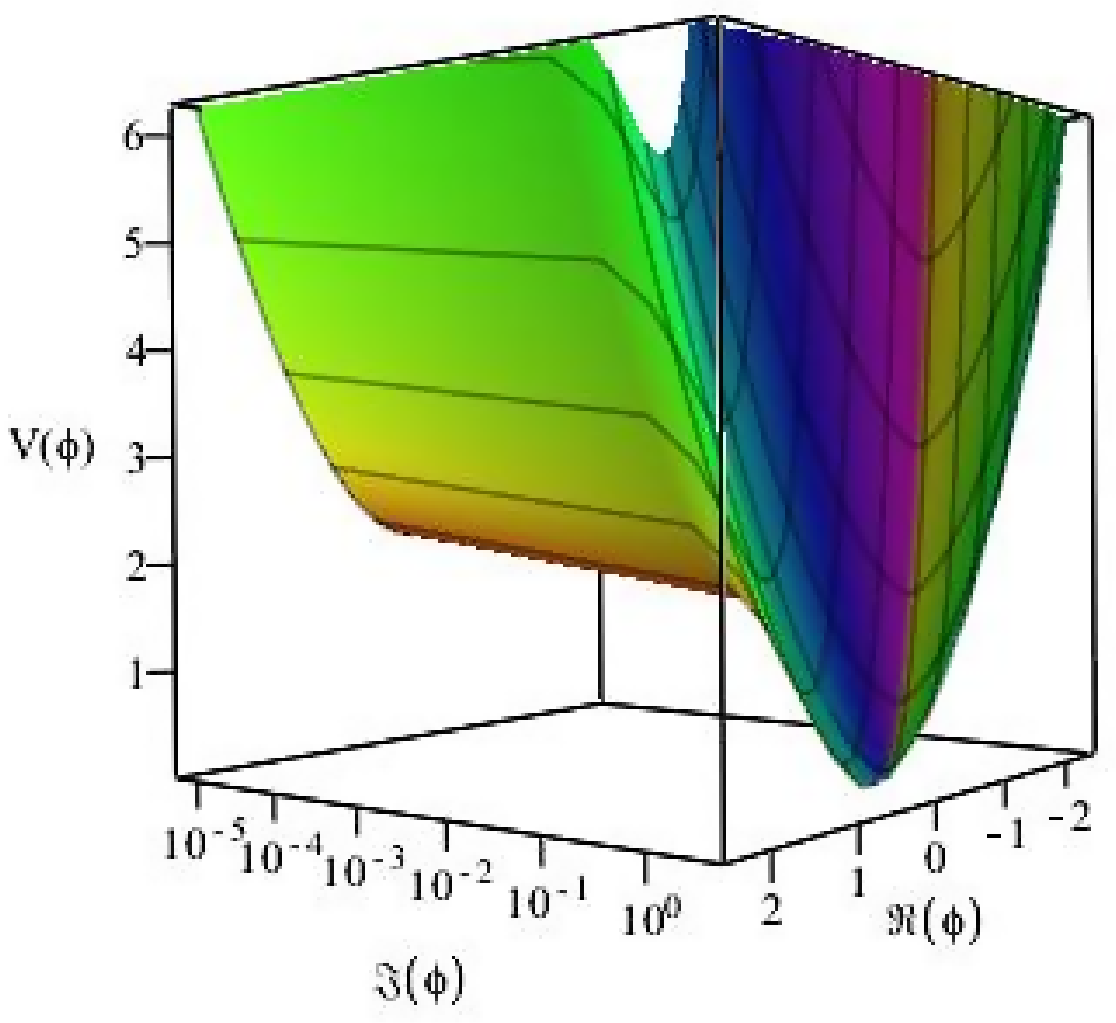}}
\caption{The 3D potential vs. both real and imaginary parts of the scalar field:
\subref{fig6a} The side face shows that the imaginary part of the scalar field is a slow roll potential, while the front face shows that the real part of the scalar field is a quadratic potential;
\subref{fig6b} The complex potential shows Higgs features.}
\label{Fig6}
\end{figure}
\begin{table}[t]
\caption{The local minima of the potentials (\ref{V1})-(\ref{V4})}
\label{T2}
\begin{tabular*}{\columnwidth}{@{\extracolsep{\fill}}lc@{}}
\hline
The potential & Local extremae at $\varphi$\\
\hline
$V_{1^{\pm}}$ & $-1.70\times 10^{-10}\pm 0.27\Im$\\
              & $-3.17\times 10^{-10}\pm 1.41\Im$ \\
\hline
$V_{2}$ & $-3.61\times 10^{-10}-1.41\Im$\\
        & $-9.66\times 10^{-11}+0.27\Im$ \\
\hline
$V_{3}$ & $0$\\
        & $-3.69\times10^{-11}\pm 1.36\Im$\\
\hline
$V_{4^{\pm}}$ & $+5.99\times 10^{-11}\pm 1.41\Im$\\
              & $-7.90\times 10^{-12}\mp 1.54\Im$\\
\hline
\end{tabular*}
\end{table}
\begin{equation}\label{V3}
V_{3}=V_{0}+\frac{\varphi^{2}}{2}-\frac{\varphi}{\sqrt{6}}\frac{\Xi^{2}-8}{\Xi}
+\frac{1}{12}\frac{\left[\Xi^{2}-8\right]^{2}}{\Xi^{2}},
\end{equation}
where $\Xi:=\left[6\sqrt{6}\varphi+2\sqrt{128+54\varphi^{2}}\right]^{1/3}$. The plot of the $V_{3}$ potential, Figure \ref{Fig5}, allows the real scalar field in the full domain ($-\infty<\varphi<\infty$) providing a quadratic-like inflation model. Interestingly, if complex scalar field is allowed, the $V_{3}$ model is quadratic at high energy. But the potential has two minima at the ground state just like Higgs potential, see Figure \ref{Fig6}\subref{fig6c}. Finally, we evaluate other two versions of the vacuum potential
\begin{eqnarray}\label{V4}
\nonumber    V_{4^{\pm}}&=&V_{0}+\frac{\varphi^{2}}{2}+\frac{\varphi}{2\sqrt{6}}\frac{\left[(\Xi^{2}-8)\pm \sqrt{3}\Im(\Xi^{2}+8)\right]}{\Xi}\\
&+&\frac{1}{48}\frac{\left[(\Xi^{2}-8)\pm \sqrt{3}\Im(\Xi^{2}+8)\right]^{2}}{\Xi^{2}},
\end{eqnarray}
the $V_{4^{\pm}}$ potentials require the scalar field to have only complex value, See Figure \ref{Fig6}\subref{fig6d}. The scalar field of the minimum potential are investigated and summarized in table \ref{T2}.
\section{Final remarks}\label{S8}
In this work, we reinvestigate the FRW universe in the absence of ordinary matter when it is governed by the $f(T)$ gravity. Fortunately, reformulation of the FRW model within $f(T)$ gravity framework does not imply vanishing of the teleparallel torsion scalar. This enabled to study the vacuum energy density in different context without imposing an artificial cosmological constant. We summarize the results of the $f(T)$ theory in the following points:\\

(\textbf{i}) The solution of the $f(T)$ field equations in the absence of matter gives an exponential scale factor which characterizes the cosmic inflation (de Sitter) models. Also, the field equations enable to construct an exact $f(T)$ gravity theory of an empty universe. However, the curvature density parameter in the non-flat space models decays to flat space limit $\Omega_{k} \rightarrow 0$, the torsion contribution in the Friedmann equations compensates exactly the missing ingredient of vacuum keeping the total density parameter to chase the unity as required by inflationary models and by cosmic and astrophysical observations as well.\\

(\textbf{ii}) The flat space model gives $\Lambda$ de Sitter universe, where the $f(T)$ acts just like the cosmological constant. The flat model is invariant under time and space translations so that the universe cannot evolve. Actually, the flat model might be useful at late cosmic evolution but early universe cannot be exactly de Sitter. On the other hand, the $f(T)$ of the non flat models have dynamical evolutions. In particular, the open universe model gives an exponentially decaying $f(T)$. Remarkably, the $f(T)$ has much larger value at early universe than its present value by $\sim 118$ orders of magnitude. This contributes directly to solve the fine-tuning problem of the cosmological constant.\\

(\textbf{iii}) We assume that the torsion vacuum is governed by perfect fluid EoS. The $f(T)$ theory can predict an initial quintessence field $\omega_{T}=-1/3$ with a de Sitter fate $\omega_{T}=-1$ in the $H>0$ spacetime regime, while The EoS parameter evolves from $-1 \rightarrow -1/3$ in the $H<0$ spacetime regime similar to the tracker field models. As is clear that the torsion vacuum fluid violates the energy conditions providing different inflationary scenarios. However, the torsion EoS shows a sudden singularity at the open universe model allowing $\omega_{T}>0$ at some early time.\\

(\textbf{iv}) Instead of using the perturbation technique, we consider the case of the propagating torsion when the tlaplon field serves as a torsion potential. This treatment enabled us to induce a scalar field sensitive to the vierbein field, i.e. the spacetime symmetry. Also, it enabled to construct an effective potential from the $f(T)$ gravity theory. This treatment enabled us to reformulate the vacuum theory in terms of a scalar field $\varphi$ in a potential $V(\varphi)$.\\

(\textbf{v}) The phase diagram plots of the induced teleparallel scalar field, its kinetic energy and effective potential vs. the evolution of the curvature density parameter timeline set some restrictions on the cosmic evolution in the non flat models. The closed universe model rejects the $\Omega_{k}\leq -1$ as it allows phantom (ghost) dark energy. The cosmic evolution is powered by $V\gg \frac{1}{2}\dot{\varphi}^{2}$ at the period $-1<\Omega<-0.696$ which derives the universe to cosmic inflation epoch. The kinetic and potential equality at $\Omega_{k}\sim -0.696$ just after the scalar field starts to decay at $\Omega_{k}=-0.704$ which provides the required condition to end the inflation. The potential goes to negative values at $\Omega_{k}<-0.473$. The scalar field decaying rate decreases as the cosmic universe goes to its flat fate. In the open universe the early cosmic acceleration is powered by vacuum potential at $\Omega\gg 1$, while the inflation ends, at $V=\frac{1}{2}\dot{\varphi}^{2}$ equality, at $\Omega_{k}=1$ which matches the sudden singularity of the torsion EoS. This indicates phase transition to radiation dominant universe as required by the end of the inflation epoch through reheating process. at $0<\Omega_{k}<1$ the universe goes to kinetic energy dominant phase which has an important impact on the large scale structure and the galaxy formation. The teleparallel scalar field shows a decaying behaviour at all time. However, its decaying rate decreases as the universe evolves towards the flat limit $\Omega_{k} \rightarrow 0$.\\

(\textbf{vi}) We evaluated six different versions of the effective potential. By considering only the real scalar field contribution we get a quadratic like inflation model. Interestingly, if we allow the complex scalar field contribution, one of the solutions gives Higgs like inflation model.\\

In conclusion, we find that the gravity formulation in the teleparallel geometry is capable to give much better descriptions of gravity. The $f(T)$ theory in the absence of matter allows the universe to evolve only in non-flat cases. Moreover, it suggests that the universe should be open hyper-sphere allowing the $f(T)$ to evolve as a decaying cosmological constant. Furthermore, the theory enforces the universe to seek the flat limit $\Omega_{k}\sim 0$ at present which matches the observations requirements without imposing the flat condition $k=0$. On the other hand, the induced tlaplon field shows remarkable effective potentials of vacuum. These potentials need more investigations to match possible energy scales of the experimental physics.
\subsection*{Acknowledgments}
This work is partially supported by the Egyptian Ministry of Scientific Research under project No. 24-2-12.
\bibliographystyle{apsrev}

\begin{thebibliography}{75}
\expandafter\ifx\csname natexlab\endcsname\relax\def\natexlab#1{#1}\fi
\expandafter\ifx\csname bibnamefont\endcsname\relax
  \def\bibnamefont#1{#1}\fi
\expandafter\ifx\csname bibfnamefont\endcsname\relax
  \def\bibfnamefont#1{#1}\fi
\expandafter\ifx\csname citenamefont\endcsname\relax
  \def\citenamefont#1{#1}\fi
\expandafter\ifx\csname url\endcsname\relax
  \def\url#1{\texttt{#1}}\fi
\expandafter\ifx\csname urlprefix\endcsname\relax\def\urlprefix{URL }\fi
\providecommand{\bibinfo}[2]{#2}
\providecommand{\eprint}[2][]{\url{#2}}

\bibitem[{\citenamefont{{Maluf} et~al.}(2002)\citenamefont{{Maluf}, {da
  Rocha-Neto}, and {Tor\'{\i}bio}}}]{M2002}
\bibinfo{author}{\bibfnamefont{J.~W.} \bibnamefont{{Maluf}}},
  \bibinfo{author}{\bibfnamefont{J.~F.} \bibnamefont{{da Rocha-Neto}}},
  \bibnamefont{and} \bibinfo{author}{\bibfnamefont{C.~K.~H.}
  \bibnamefont{{Tor\'{\i}bio}}, \bibfnamefont{T.~M.~and}},
  \bibinfo{journal}{Phys. Rev. D} \textbf{\bibinfo{volume}{65}},
  \bibinfo{eid}{124001} (\bibinfo{year}{2002}), \eprint{gr-qc/0204035}.

\bibitem[{\citenamefont{{Li} et~al.}(2011)\citenamefont{{Li}, {Sotiriou}, and
  {Barrow}}}]{1010.1041}
\bibinfo{author}{\bibfnamefont{B.}~\bibnamefont{{Li}}},
  \bibinfo{author}{\bibfnamefont{T.~P.} \bibnamefont{{Sotiriou}}},
  \bibnamefont{and} \bibinfo{author}{\bibfnamefont{J.~D.}
  \bibnamefont{{Barrow}}}, \bibinfo{journal}{Phys. Rev. D}
  \textbf{\bibinfo{volume}{83}}, \bibinfo{eid}{064035} (\bibinfo{year}{2011}),
  \eprint{1010.1041}.

\bibitem[{\citenamefont{{Sotiriou} et~al.}(2011)\citenamefont{{Sotiriou}, {Li},
  and {Barrow}}}]{1012.4039}
\bibinfo{author}{\bibfnamefont{T.~P.} \bibnamefont{{Sotiriou}}},
  \bibinfo{author}{\bibfnamefont{B.}~\bibnamefont{{Li}}}, \bibnamefont{and}
  \bibinfo{author}{\bibfnamefont{J.~D.} \bibnamefont{{Barrow}}},
  \bibinfo{journal}{Phys. Rev. D} \textbf{\bibinfo{volume}{83}},
  \bibinfo{eid}{104030} (\bibinfo{year}{2011}), \eprint{1012.4039}.

\bibitem[{\citenamefont{{Ferraro} and {Fiorini}}(2007)}]{FF07}
\bibinfo{author}{\bibfnamefont{R.}~\bibnamefont{{Ferraro}}} \bibnamefont{and}
  \bibinfo{author}{\bibfnamefont{F.}~\bibnamefont{{Fiorini}}},
  \bibinfo{journal}{Phys. Rev. D} \textbf{\bibinfo{volume}{75}},
  \bibinfo{eid}{084031} (\bibinfo{year}{2007}), \eprint{gr-qc/0610067}.

\bibitem[{\citenamefont{{Ferraro} and {Fiorini}}(2008)}]{FF08}
\bibinfo{author}{\bibfnamefont{R.}~\bibnamefont{{Ferraro}}} \bibnamefont{and}
  \bibinfo{author}{\bibfnamefont{F.}~\bibnamefont{{Fiorini}}},
  \bibinfo{journal}{Phys. Rev. D} \textbf{\bibinfo{volume}{78}},
  \bibinfo{eid}{124019} (\bibinfo{year}{2008}), \eprint{gr-qc/0812.1981}.

\bibitem[{\citenamefont{{Bengochea} and {Ferraro}}(2009)}]{BF09}
\bibinfo{author}{\bibfnamefont{G.~R.} \bibnamefont{{Bengochea}}}
  \bibnamefont{and}
  \bibinfo{author}{\bibfnamefont{R.}~\bibnamefont{{Ferraro}}},
  \bibinfo{journal}{Phys. Rev. D} \textbf{\bibinfo{volume}{79}},
  \bibinfo{eid}{124019} (\bibinfo{year}{2009}), \eprint{0812.1205}.

\bibitem[{\citenamefont{{Linder}}(2010)}]{L10}
\bibinfo{author}{\bibfnamefont{E.~V.} \bibnamefont{{Linder}}},
  \bibinfo{journal}{Phys. Rev. D} \textbf{\bibinfo{volume}{81}},
  \bibinfo{eid}{127301} (\bibinfo{year}{2010}), \eprint{1005.3039}.

\bibitem[{\citenamefont{{Bamba} et~al.}(2010)\citenamefont{{Bamba}, {Geng}, and
  {Lee}}}]{1008.4036}
\bibinfo{author}{\bibfnamefont{K.}~\bibnamefont{{Bamba}}},
  \bibinfo{author}{\bibfnamefont{C.-Q.} \bibnamefont{{Geng}}},
  \bibnamefont{and} \bibinfo{author}{\bibfnamefont{C.-C.} \bibnamefont{{Lee}}},
  \bibinfo{journal}{ArXiv e-prints}  (\bibinfo{year}{2010}),
  \eprint{1008.4036}.

\bibitem[{\citenamefont{{Bamba} et~al.}(2011)\citenamefont{{Bamba}, {Geng},
  {Lee}, and {Luo}}}]{1011.0508}
\bibinfo{author}{\bibfnamefont{K.}~\bibnamefont{{Bamba}}},
  \bibinfo{author}{\bibfnamefont{C.-Q.} \bibnamefont{{Geng}}},
  \bibinfo{author}{\bibfnamefont{C.-C.} \bibnamefont{{Lee}}}, \bibnamefont{and}
  \bibinfo{author}{\bibfnamefont{L.-W.} \bibnamefont{{Luo}}},
  \bibinfo{journal}{JCAP} \textbf{\bibinfo{volume}{1}}, \bibinfo{eid}{021}
  (\bibinfo{year}{2011}), \eprint{1011.0508}.

\bibitem[{\citenamefont{{Yang}}(2011)}]{Y2011}
\bibinfo{author}{\bibfnamefont{R.-J.} \bibnamefont{{Yang}}},
  \bibinfo{journal}{EPL (Europhysics Letters)} \textbf{\bibinfo{volume}{93}},
  \bibinfo{pages}{60001} (\bibinfo{year}{2011}), \eprint{1010.1376}.

\bibitem[{\citenamefont{{Cai} et~al.}(2011)\citenamefont{{Cai}, {Chen}, {Dent},
  {Dutta}, and {Saridakis}}}]{CCDDS11}
\bibinfo{author}{\bibfnamefont{Y.-F.} \bibnamefont{{Cai}}},
  \bibinfo{author}{\bibfnamefont{S.-H.} \bibnamefont{{Chen}}},
  \bibinfo{author}{\bibfnamefont{J.~B.} \bibnamefont{{Dent}}},
  \bibinfo{author}{\bibfnamefont{S.}~\bibnamefont{{Dutta}}}, \bibnamefont{and}
  \bibinfo{author}{\bibfnamefont{E.~N.} \bibnamefont{{Saridakis}}},
  \bibinfo{journal}{Classical and Quantum Gravity}
  \textbf{\bibinfo{volume}{28}}, \bibinfo{eid}{215011} (\bibinfo{year}{2011}),
  \eprint{1104.4349}.

\bibitem[{\citenamefont{{Ferraro} and {Fiorini}}(2011{\natexlab{a}})}]{FF011}
\bibinfo{author}{\bibfnamefont{R.}~\bibnamefont{{Ferraro}}} \bibnamefont{and}
  \bibinfo{author}{\bibfnamefont{F.}~\bibnamefont{{Fiorini}}},
  \bibinfo{journal}{Phys. Rev. D} \textbf{\bibinfo{volume}{84}},
  \bibinfo{eid}{083518} (\bibinfo{year}{2011}{\natexlab{a}}),
  \eprint{1109.4209}.

\bibitem[{\citenamefont{{Ferraro} and {Fiorini}}(2011{\natexlab{b}})}]{FF11}
\bibinfo{author}{\bibfnamefont{R.}~\bibnamefont{{Ferraro}}} \bibnamefont{and}
  \bibinfo{author}{\bibfnamefont{F.}~\bibnamefont{{Fiorini}}},
  \bibinfo{journal}{Phys. Lett. B} \textbf{\bibinfo{volume}{702}},
  \bibinfo{pages}{75} (\bibinfo{year}{2011}{\natexlab{b}}), \eprint{1103.0824}.

\bibitem[{\citenamefont{{Iorio} and {Saridakis}}(2012)}]{IS12}
\bibinfo{author}{\bibfnamefont{L.}~\bibnamefont{{Iorio}}} \bibnamefont{and}
  \bibinfo{author}{\bibfnamefont{E.~N.} \bibnamefont{{Saridakis}}},
  \bibinfo{journal}{Mon. Not. R. Astron. Soc.} \textbf{\bibinfo{volume}{427}},
  \bibinfo{pages}{1555} (\bibinfo{year}{2012}), \eprint{1203.5781}.

\bibitem[{\citenamefont{{Capozziello} et~al.}(2013)\citenamefont{{Capozziello},
  {Gonz{\'a}lez}, {Saridakis}, and {V{\'a}squez}}}]{CGS13}
\bibinfo{author}{\bibfnamefont{S.}~\bibnamefont{{Capozziello}}},
  \bibinfo{author}{\bibfnamefont{P.~A.} \bibnamefont{{Gonz{\'a}lez}}},
  \bibinfo{author}{\bibfnamefont{E.~N.} \bibnamefont{{Saridakis}}},
  \bibnamefont{and}
  \bibinfo{author}{\bibfnamefont{Y.}~\bibnamefont{{V{\'a}squez}}},
  \bibinfo{journal}{JHEP} \textbf{\bibinfo{volume}{2}}, \bibinfo{eid}{39}
  (\bibinfo{year}{2013}), \eprint{1210.1098}.

\bibitem[{\citenamefont{{Shirafuji} et~al.}(1996)\citenamefont{{Shirafuji},
  {Nashed}, and {Kobayashi}}}]{Nashed:1996}
\bibinfo{author}{\bibfnamefont{T.}~\bibnamefont{{Shirafuji}}},
  \bibinfo{author}{\bibfnamefont{G.~G.} \bibnamefont{{Nashed}}},
  \bibnamefont{and}
  \bibinfo{author}{\bibfnamefont{Y.}~\bibnamefont{{Kobayashi}}},
  \bibinfo{journal}{Progress of Theoretical Physics}
  \textbf{\bibinfo{volume}{96}}, \bibinfo{pages}{933} (\bibinfo{year}{1996}),
  \eprint{gr-qc/9609060}.

\bibitem[{\citenamefont{{Shirafuji} and {Nashed}}(1997)}]{Nashed:1997}
\bibinfo{author}{\bibfnamefont{T.}~\bibnamefont{{Shirafuji}}} \bibnamefont{and}
  \bibinfo{author}{\bibfnamefont{G.~G.} \bibnamefont{{Nashed}}},
  \bibinfo{journal}{Progress of Theoretical Physics}
  \textbf{\bibinfo{volume}{98}}, \bibinfo{pages}{1355} (\bibinfo{year}{1997}),
  \eprint{gr-qc/9711010}.

\bibitem[{\citenamefont{{Nashed}}(2002)}]{Nashed:2002}
\bibinfo{author}{\bibfnamefont{G.~G.~L.} \bibnamefont{{Nashed}}},
  \bibinfo{journal}{Nuovo Cimento B Serie} \textbf{\bibinfo{volume}{117}},
  \bibinfo{pages}{521} (\bibinfo{year}{2002}), \eprint{gr-qc/0109017}.

\bibitem[{\citenamefont{{Nashed}}(2006)}]{Nashed:2006}
\bibinfo{author}{\bibfnamefont{G.~G.~L.} \bibnamefont{{Nashed}}},
  \bibinfo{journal}{International Journal of Modern Physics A}
  \textbf{\bibinfo{volume}{21}}, \bibinfo{pages}{3181} (\bibinfo{year}{2006}),
  \eprint{gr-qc/0501002}.

\bibitem[{\citenamefont{{Nashed}}(2007)}]{Nashed:2007}
\bibinfo{author}{\bibfnamefont{G.~G.~L.} \bibnamefont{{Nashed}}},
  \bibinfo{journal}{European Physical Journal C} \textbf{\bibinfo{volume}{49}},
  \bibinfo{pages}{851} (\bibinfo{year}{2007}), \eprint{0706.0260}.

\bibitem[{\citenamefont{Nashed}(2012)}]{Nashed:2012}
\bibinfo{author}{\bibfnamefont{G.~G.~L.} \bibnamefont{Nashed}},
  \bibinfo{journal}{Chinese Physics Letters} \textbf{\bibinfo{volume}{29}},
  \bibinfo{pages}{050402} (\bibinfo{year}{2012}).

\bibitem[{\citenamefont{{Nashed}}(2014)}]{Nashed3}
\bibinfo{author}{\bibfnamefont{G.~G.~L.} \bibnamefont{{Nashed}}},
  \bibinfo{journal}{EPL (Europhysics Letters)} \textbf{\bibinfo{volume}{105}},
  \bibinfo{eid}{10001} (\bibinfo{year}{2014}), \eprint{1501.00974}.

\bibitem[{\citenamefont{{Nashed}}(2015)}]{Nashed5}
\bibinfo{author}{\bibfnamefont{G.~G.~L.} \bibnamefont{{Nashed}}},
  \bibinfo{journal}{International Journal of Modern Physics D}
  \textbf{\bibinfo{volume}{24}}, \bibinfo{eid}{1550007} (\bibinfo{year}{2015}).

\bibitem[{\citenamefont{{Rodrigues} et~al.}(2013)\citenamefont{{Rodrigues},
  {Houndjo}, {Tossa}, {Momeni}, and {Myrzakulov}}}]{RHTMM13}
\bibinfo{author}{\bibfnamefont{M.~E.} \bibnamefont{{Rodrigues}}},
  \bibinfo{author}{\bibfnamefont{M.~J.~S.} \bibnamefont{{Houndjo}}},
  \bibinfo{author}{\bibfnamefont{J.}~\bibnamefont{{Tossa}}},
  \bibinfo{author}{\bibfnamefont{D.}~\bibnamefont{{Momeni}}}, \bibnamefont{and}
  \bibinfo{author}{\bibfnamefont{R.}~\bibnamefont{{Myrzakulov}}},
  \bibinfo{journal}{JCAP} \textbf{\bibinfo{volume}{11}}, \bibinfo{eid}{024}
  (\bibinfo{year}{2013}), \eprint{1306.2280}.

\bibitem[{\citenamefont{{Bejarano} et~al.}(2015)\citenamefont{{Bejarano},
  {Ferraro}, and {Guzm{\'a}n}}}]{BFG15}
\bibinfo{author}{\bibfnamefont{C.}~\bibnamefont{{Bejarano}}},
  \bibinfo{author}{\bibfnamefont{R.}~\bibnamefont{{Ferraro}}},
  \bibnamefont{and} \bibinfo{author}{\bibfnamefont{M.~J.}
  \bibnamefont{{Guzm{\'a}n}}}, \bibinfo{journal}{Eur. Phys. J. C}
  \textbf{\bibinfo{volume}{75}}, \bibinfo{pages}{77} (\bibinfo{year}{2015}),
  \eprint{1412.0641}.

\bibitem[{\citenamefont{{Iorio} et~al.}(2015)\citenamefont{{Iorio},
  {Radicella}, and {Ruggiero}}}]{IRR2015}
\bibinfo{author}{\bibfnamefont{L.}~\bibnamefont{{Iorio}}},
  \bibinfo{author}{\bibfnamefont{N.}~\bibnamefont{{Radicella}}},
  \bibnamefont{and} \bibinfo{author}{\bibfnamefont{M.~L.}
  \bibnamefont{{Ruggiero}}}, \bibinfo{journal}{JCAP}
  \textbf{\bibinfo{volume}{8}}, \bibinfo{eid}{021} (\bibinfo{year}{2015}),
  \eprint{1505.06996}.

\bibitem[{\citenamefont{{Karami} et~al.}(2013)\citenamefont{{Karami},
  {Abdolmaleki}, {Asadzadeh}, and {Safari}}}]{KAAS13}
\bibinfo{author}{\bibfnamefont{K.}~\bibnamefont{{Karami}}},
  \bibinfo{author}{\bibfnamefont{A.}~\bibnamefont{{Abdolmaleki}}},
  \bibinfo{author}{\bibfnamefont{S.}~\bibnamefont{{Asadzadeh}}},
  \bibnamefont{and} \bibinfo{author}{\bibfnamefont{Z.}~\bibnamefont{{Safari}}},
  \bibinfo{journal}{European Physical Journal C} \textbf{\bibinfo{volume}{73}},
  \bibinfo{eid}{2565} (\bibinfo{year}{2013}), \eprint{1202.2278}.

\bibitem[{\citenamefont{{Bamba} et~al.}(2012)\citenamefont{{Bamba},
  {Capozziello}, {Nojiri}, and {Odintsov}}}]{1205.3421}
\bibinfo{author}{\bibfnamefont{K.}~\bibnamefont{{Bamba}}},
  \bibinfo{author}{\bibfnamefont{S.}~\bibnamefont{{Capozziello}}},
  \bibinfo{author}{\bibfnamefont{S.}~\bibnamefont{{Nojiri}}}, \bibnamefont{and}
  \bibinfo{author}{\bibfnamefont{S.~D.} \bibnamefont{{Odintsov}}},
  \bibinfo{journal}{Astrophysics and Space Science}
  \textbf{\bibinfo{volume}{342}}, \bibinfo{pages}{155} (\bibinfo{year}{2012}),
  \eprint{1205.3421}.

\bibitem[{\citenamefont{{Bamba} et~al.}(2014)\citenamefont{{Bamba}, {Nojiri},
  and {Odintsov}}}]{BNO14}
\bibinfo{author}{\bibfnamefont{K.}~\bibnamefont{{Bamba}}},
  \bibinfo{author}{\bibfnamefont{S.}~\bibnamefont{{Nojiri}}}, \bibnamefont{and}
  \bibinfo{author}{\bibfnamefont{S.~D.} \bibnamefont{{Odintsov}}},
  \bibinfo{journal}{Phys. Lett. B} \textbf{\bibinfo{volume}{731}},
  \bibinfo{pages}{257} (\bibinfo{year}{2014}), \eprint{1401.7378}.

\bibitem[{\citenamefont{{Bamba} and {Odintsov}}(2014)}]{BO14}
\bibinfo{author}{\bibfnamefont{K.}~\bibnamefont{{Bamba}}} \bibnamefont{and}
  \bibinfo{author}{\bibfnamefont{S.~D.} \bibnamefont{{Odintsov}}},
  \bibinfo{journal}{ArXiv: 1402.7114}  (\bibinfo{year}{2014}),
  \eprint{1402.7114}.

\bibitem[{\citenamefont{{Jamil} et~al.}(2014)\citenamefont{{Jamil}, {Momeni},
  and {Myrzakulov}}}]{JMM14}
\bibinfo{author}{\bibfnamefont{M.}~\bibnamefont{{Jamil}}},
  \bibinfo{author}{\bibfnamefont{D.}~\bibnamefont{{Momeni}}}, \bibnamefont{and}
  \bibinfo{author}{\bibfnamefont{R.}~\bibnamefont{{Myrzakulov}}},
  \bibinfo{journal}{International Journal of Theoretical Physics}
  (\bibinfo{year}{2014}), \eprint{1309.3269}.

\bibitem[{\citenamefont{{Harko} et~al.}(2014)\citenamefont{{Harko}, {Lobo},
  {Otalora}, and {Saridakis}}}]{HLOS14}
\bibinfo{author}{\bibfnamefont{T.}~\bibnamefont{{Harko}}},
  \bibinfo{author}{\bibfnamefont{F.~S.~N.} \bibnamefont{{Lobo}}},
  \bibinfo{author}{\bibfnamefont{G.}~\bibnamefont{{Otalora}}},
  \bibnamefont{and} \bibinfo{author}{\bibfnamefont{E.~N.}
  \bibnamefont{{Saridakis}}}, \bibinfo{journal}{Phys. Rev. D}
  \textbf{\bibinfo{volume}{89}}, \bibinfo{eid}{124036} (\bibinfo{year}{2014}),
  \eprint{1404.6212}.

\bibitem[{\citenamefont{{Nashed} and {El Hanafy}}(2014)}]{NH14}
\bibinfo{author}{\bibfnamefont{G.~G.~L.} \bibnamefont{{Nashed}}}
  \bibnamefont{and} \bibinfo{author}{\bibfnamefont{W.}~\bibnamefont{{El
  Hanafy}}}, \bibinfo{journal}{Eur. Phys. J. C} \textbf{\bibinfo{volume}{74}},
  \bibinfo{pages}{3099} (\bibinfo{year}{2014}), \bibinfo{note}{arXiv:
  1403.0913}, \eprint{gr-qc/1403.0913}.

\bibitem[{\citenamefont{{Wu} et~al.}(2015)\citenamefont{{Wu}, {Chen}, {Wang},
  and {Wei}}}]{WCWW2015}
\bibinfo{author}{\bibfnamefont{Y.}~\bibnamefont{{Wu}}},
  \bibinfo{author}{\bibfnamefont{Z.-C.} \bibnamefont{{Chen}}},
  \bibinfo{author}{\bibfnamefont{J.-X.} \bibnamefont{{Wang}}},
  \bibnamefont{and} \bibinfo{author}{\bibfnamefont{H.}~\bibnamefont{{Wei}}},
  \bibinfo{journal}{Communications in Theoretical Physics}
  \textbf{\bibinfo{volume}{63}}, \bibinfo{eid}{701} (\bibinfo{year}{2015}),
  \eprint{1503.05281}.

\bibitem[{\citenamefont{{Junior} et~al.}(2015)\citenamefont{{Junior},
  {Rodrigues}, and {Houndjo}}}]{1503.07427}
\bibinfo{author}{\bibfnamefont{E.~L.~B.} \bibnamefont{{Junior}}},
  \bibinfo{author}{\bibfnamefont{M.~E.} \bibnamefont{{Rodrigues}}},
  \bibnamefont{and} \bibinfo{author}{\bibfnamefont{M.~J.~S.}
  \bibnamefont{{Houndjo}}}, \bibinfo{journal}{JCAP}
  \textbf{\bibinfo{volume}{6}}, \bibinfo{eid}{037} (\bibinfo{year}{2015}),
  \eprint{1503.07427}.

\bibitem[{\citenamefont{{El Hanafy} and {Nashed}}(2015)}]{HN15}
\bibinfo{author}{\bibfnamefont{W.}~\bibnamefont{{El Hanafy}}} \bibnamefont{and}
  \bibinfo{author}{\bibfnamefont{G.~G.~L.} \bibnamefont{{Nashed}}},
  \bibinfo{journal}{Eur. Phys. J. C} \textbf{\bibinfo{volume}{75}},
  \bibinfo{eid}{279} (\bibinfo{year}{2015}), \eprint{hep-th/1409.7199}.

\bibitem[{\citenamefont{{Rezazadeh} et~al.}(2015)\citenamefont{{Rezazadeh},
  {Abdolmaleki}, and {Karami}}}]{RAK2015}
\bibinfo{author}{\bibfnamefont{K.}~\bibnamefont{{Rezazadeh}}},
  \bibinfo{author}{\bibfnamefont{A.}~\bibnamefont{{Abdolmaleki}}},
  \bibnamefont{and} \bibinfo{author}{\bibfnamefont{K.}~\bibnamefont{{Karami}}},
  \bibinfo{journal}{ArXiv e-prints}  (\bibinfo{year}{2015}),
  \eprint{1509.08769}.

\bibitem[{\citenamefont{{Jennen} and {Pereira}}(2016)}]{JP2016}
\bibinfo{author}{\bibfnamefont{H.}~\bibnamefont{{Jennen}}} \bibnamefont{and}
  \bibinfo{author}{\bibfnamefont{J.~G.} \bibnamefont{{Pereira}}},
  \bibinfo{journal}{Physics of the Dark Universe}
  \textbf{\bibinfo{volume}{11}}, \bibinfo{pages}{49} (\bibinfo{year}{2016}),
  \eprint{1506.02012}.

\bibitem[{\citenamefont{{Cai} et~al.}(2014)\citenamefont{{Cai}, {Quintin},
  {Saridakis}, and {Wilson-Ewing}}}]{CQSW14}
\bibinfo{author}{\bibfnamefont{Y.-F.} \bibnamefont{{Cai}}},
  \bibinfo{author}{\bibfnamefont{J.}~\bibnamefont{{Quintin}}},
  \bibinfo{author}{\bibfnamefont{E.~N.} \bibnamefont{{Saridakis}}},
  \bibnamefont{and}
  \bibinfo{author}{\bibfnamefont{E.}~\bibnamefont{{Wilson-Ewing}}},
  \bibinfo{journal}{JCAP} \textbf{\bibinfo{volume}{7}}, \bibinfo{eid}{033}
  (\bibinfo{year}{2014}), \eprint{1404.4364}.

\bibitem[{\citenamefont{Amorós et~al.}(2013)\citenamefont{Amorós, de~Haro, and
  Odintsov}}]{bounce3}
\bibinfo{author}{\bibfnamefont{J.}~\bibnamefont{Amorós}},
  \bibinfo{author}{\bibfnamefont{J.}~\bibnamefont{de~Haro}}, \bibnamefont{and}
  \bibinfo{author}{\bibfnamefont{S.~D.} \bibnamefont{Odintsov}},
  \bibinfo{journal}{Physical Review D} \textbf{\bibinfo{volume}{87}},
  \bibinfo{pages}{104037} (\bibinfo{year}{2013}), \eprint{1305.2344}.

\bibitem[{\citenamefont{{Bamba} et~al.}(2016)\citenamefont{{Bamba}, {Nashed},
  {El Hanafy}, and {Ibrahim}}}]{BNEI:2016}
\bibinfo{author}{\bibfnamefont{K.}~\bibnamefont{{Bamba}}},
  \bibinfo{author}{\bibfnamefont{G.~G.~L.} \bibnamefont{{Nashed}}},
  \bibinfo{author}{\bibfnamefont{W.}~\bibnamefont{{El Hanafy}}},
  \bibnamefont{and} \bibinfo{author}{\bibfnamefont{S.~K.}
  \bibnamefont{{Ibrahim}}}, \bibinfo{journal}{ArXiv e-prints}
  (\bibinfo{year}{2016}), \eprint{1604.07604}.

\bibitem[{\citenamefont{{Kofinas} and {Saridakis}}(2014{\natexlab{a}})}]{KS114}
\bibinfo{author}{\bibfnamefont{G.}~\bibnamefont{{Kofinas}}} \bibnamefont{and}
  \bibinfo{author}{\bibfnamefont{E.~N.} \bibnamefont{{Saridakis}}},
  \bibinfo{journal}{Phys. Rev. D} \textbf{\bibinfo{volume}{90}},
  \bibinfo{eid}{084044} (\bibinfo{year}{2014}{\natexlab{a}}),
  \eprint{1404.2249}.

\bibitem[{\citenamefont{{Kofinas} and {Saridakis}}(2014{\natexlab{b}})}]{KS214}
\bibinfo{author}{\bibfnamefont{G.}~\bibnamefont{{Kofinas}}} \bibnamefont{and}
  \bibinfo{author}{\bibfnamefont{E.~N.} \bibnamefont{{Saridakis}}},
  \bibinfo{journal}{Phys. Rev. D} \textbf{\bibinfo{volume}{90}},
  \bibinfo{eid}{084045} (\bibinfo{year}{2014}{\natexlab{b}}),
  \eprint{1408.0107}.

\bibitem[{\citenamefont{{Kofinas} et~al.}(2014)\citenamefont{{Kofinas}, {Leon},
  and {Saridakis}}}]{KS314}
\bibinfo{author}{\bibfnamefont{G.}~\bibnamefont{{Kofinas}}},
  \bibinfo{author}{\bibfnamefont{G.}~\bibnamefont{{Leon}}}, \bibnamefont{and}
  \bibinfo{author}{\bibfnamefont{E.~N.} \bibnamefont{{Saridakis}}},
  \bibinfo{journal}{Classical and Quantum Gravity}
  \textbf{\bibinfo{volume}{31}}, \bibinfo{eid}{175011} (\bibinfo{year}{2014}),
  \eprint{1404.7100}.

\bibitem[{\citenamefont{{Riess} et~al.}(1998)\citenamefont{{Riess},
  {Filippenko}, {Challis}, {Clocchiatti}, {Diercks}, {Garnavich}, {Gilliland},
  {Hogan}, {Jha}, {Kirshner} et~al.}}]{SN98}
\bibinfo{author}{\bibfnamefont{A.~G.} \bibnamefont{{Riess}}},
  \bibinfo{author}{\bibfnamefont{A.~V.} \bibnamefont{{Filippenko}}},
  \bibinfo{author}{\bibfnamefont{P.}~\bibnamefont{{Challis}}},
  \bibinfo{author}{\bibfnamefont{A.}~\bibnamefont{{Clocchiatti}}},
  \bibinfo{author}{\bibfnamefont{A.}~\bibnamefont{{Diercks}}},
  \bibinfo{author}{\bibfnamefont{P.~M.} \bibnamefont{{Garnavich}}},
  \bibinfo{author}{\bibfnamefont{R.~L.} \bibnamefont{{Gilliland}}},
  \bibinfo{author}{\bibfnamefont{C.~J.} \bibnamefont{{Hogan}}},
  \bibinfo{author}{\bibfnamefont{S.}~\bibnamefont{{Jha}}},
  \bibinfo{author}{\bibfnamefont{R.~P.} \bibnamefont{{Kirshner}}},
  \bibnamefont{et~al.}, \bibinfo{journal}{Astron.~J.}
  \textbf{\bibinfo{volume}{116}}, \bibinfo{pages}{1009} (\bibinfo{year}{1998}),
  \eprint{arXiv:astro-ph/9805201}.

\bibitem[{\citenamefont{{Liddle}}(2003)}]{Liddle:2003}
\bibinfo{author}{\bibfnamefont{A.}~\bibnamefont{{Liddle}}},
  \emph{\bibinfo{title}{{An Introduction to Modern Cosmology, Second Edition}}}
  (\bibinfo{year}{2003}).

\bibitem[{\citenamefont{{Rostami} and {Jalalzadeh}}(2015)}]{RJ2015}
\bibinfo{author}{\bibfnamefont{T.}~\bibnamefont{{Rostami}}} \bibnamefont{and}
  \bibinfo{author}{\bibfnamefont{S.}~\bibnamefont{{Jalalzadeh}}},
  \bibinfo{journal}{Physics of the Dark Universe} \textbf{\bibinfo{volume}{9}},
  \bibinfo{pages}{31} (\bibinfo{year}{2015}), \eprint{1510.02068}.

\bibitem[{\citenamefont{{Wanas}}(2012)}]{W2012}
\bibinfo{author}{\bibfnamefont{M.~I.} \bibnamefont{{Wanas}}},
  \bibinfo{journal}{Adv. High Energy Phys.} \textbf{\bibinfo{volume}{2012}},
  \bibinfo{pages}{Article ID 752613, 10 pages} (\bibinfo{year}{2012}).

\bibitem[{\citenamefont{Aldrovandi and Pereira}(2013)}]{AP2013}
\bibinfo{author}{\bibfnamefont{R.}~\bibnamefont{Aldrovandi}} \bibnamefont{and}
  \bibinfo{author}{\bibfnamefont{J.~G.} \bibnamefont{Pereira}},
  \emph{\bibinfo{title}{Teleparallel Gravity: An Introduction}}, vol.
  \bibinfo{volume}{173} of \emph{\bibinfo{series}{Fundamental Theories of
  Physics}} (\bibinfo{publisher}{Springer}, \bibinfo{address}{Springer
  Dordrecht Heidelberg New York London}, \bibinfo{year}{2013}), ISBN
  \bibinfo{isbn}{978-94-007-5142-2}.

\bibitem[{\citenamefont{{Weitzenb\"{o}ck}}(1923)}]{Wr}
\bibinfo{author}{\bibfnamefont{R.}~\bibnamefont{{Weitzenb\"{o}ck}}},
  \emph{\bibinfo{title}{Invariance Theorie}} (\bibinfo{publisher}{Gronin-gen},
  \bibinfo{year}{1923}).

\bibitem[{\citenamefont{{Mikhail} and {Wanas}}(1977)}]{MW77}
\bibinfo{author}{\bibfnamefont{F.~I.} \bibnamefont{{Mikhail}}}
  \bibnamefont{and} \bibinfo{author}{\bibfnamefont{M.~I.}
  \bibnamefont{{Wanas}}}, \bibinfo{journal}{Royal Society of London Proceedings
  Series A} \textbf{\bibinfo{volume}{356}}, \bibinfo{pages}{471}
  (\bibinfo{year}{1977}).

\bibitem[{\citenamefont{{M{\o}ller}}(1978)}]{M78}
\bibinfo{author}{\bibfnamefont{C.}~\bibnamefont{{M{\o}ller}}},
  \bibinfo{journal}{K. Dan. Vidensk. Selsk. Mat. Fys. Skr.}
  \textbf{\bibinfo{volume}{89}} (\bibinfo{year}{1978}).

\bibitem[{\citenamefont{{Hayashi} and {Shirafuji}}(1979)}]{HS79}
\bibinfo{author}{\bibfnamefont{K.}~\bibnamefont{{Hayashi}}} \bibnamefont{and}
  \bibinfo{author}{\bibfnamefont{T.}~\bibnamefont{{Shirafuji}}},
  \bibinfo{journal}{Physical Review D} \textbf{\bibinfo{volume}{19}},
  \bibinfo{pages}{3524} (\bibinfo{year}{1979}).

\bibitem[{\citenamefont{{Youssef} and {Sid-Ahmed}}(2007)}]{NS07}
\bibinfo{author}{\bibfnamefont{N.~L.} \bibnamefont{{Youssef}}}
  \bibnamefont{and} \bibinfo{author}{\bibfnamefont{A.~M.}
  \bibnamefont{{Sid-Ahmed}}}, \bibinfo{journal}{Reports on Mathematical
  Physics} \textbf{\bibinfo{volume}{60}}, \bibinfo{pages}{39}
  (\bibinfo{year}{2007}), \eprint{0604111}.

\bibitem[{\citenamefont{{Youssef} and {Elsayed}}(2013)}]{NS13}
\bibinfo{author}{\bibfnamefont{N.~L.} \bibnamefont{{Youssef}}}
  \bibnamefont{and} \bibinfo{author}{\bibfnamefont{W.~A.}
  \bibnamefont{{Elsayed}}}, \bibinfo{journal}{Reports on Mathematical Physics}
  \textbf{\bibinfo{volume}{72}}, \bibinfo{pages}{1} (\bibinfo{year}{2013}),
  \eprint{1209.1379}.

\bibitem[{\citenamefont{{Wanas}}(2009)}]{W09}
\bibinfo{author}{\bibfnamefont{M.~I.} \bibnamefont{{Wanas}}},
  \bibinfo{journal}{Modern Physics Letters A} \textbf{\bibinfo{volume}{24}},
  \bibinfo{pages}{1749} (\bibinfo{year}{2009}), \eprint{0801.1132}.

\bibitem[{\citenamefont{{Wanas} and {Kamal}}(2011)}]{WK11}
\bibinfo{author}{\bibfnamefont{M.~I.} \bibnamefont{{Wanas}}} \bibnamefont{and}
  \bibinfo{author}{\bibfnamefont{M.~M.} \bibnamefont{{Kamal}}},
  \bibinfo{journal}{Modern Physics Letters A} \textbf{\bibinfo{volume}{26}},
  \bibinfo{pages}{2065} (\bibinfo{year}{2011}), \eprint{1103.4121}.

\bibitem[{\citenamefont{{Youssef} et~al.}(2013)\citenamefont{{Youssef},
  {Sid-Ahmed}, and {Taha}}}]{YST12}
\bibinfo{author}{\bibfnamefont{N.~L.} \bibnamefont{{Youssef}}},
  \bibinfo{author}{\bibfnamefont{A.~M.} \bibnamefont{{Sid-Ahmed}}},
  \bibnamefont{and} \bibinfo{author}{\bibfnamefont{E.~H.}
  \bibnamefont{{Taha}}}, \bibinfo{journal}{Int. J. Geom. Meth. Mod. Phys.}
  \textbf{\bibinfo{volume}{10}}, \bibinfo{pages}{1350029}
  (\bibinfo{year}{2013}), \eprint{1206.4505}.

\bibitem[{\citenamefont{{Wanas}}(1986)}]{Wanas86}
\bibinfo{author}{\bibfnamefont{M.~I.} \bibnamefont{{Wanas}}},
  \bibinfo{journal}{Astrophysics \& Space Science}
  \textbf{\bibinfo{volume}{127}}, \bibinfo{pages}{21} (\bibinfo{year}{1986}).

\bibitem[{\citenamefont{{Wanas} and {Ammar}}(2013)}]{WA13}
\bibinfo{author}{\bibfnamefont{M.}~\bibnamefont{{Wanas}}} \bibnamefont{and}
  \bibinfo{author}{\bibfnamefont{S.}~\bibnamefont{{Ammar}}},
  \bibinfo{journal}{Open Physics} \textbf{\bibinfo{volume}{11}},
  \bibinfo{pages}{936} (\bibinfo{year}{2013}).

\bibitem[{\citenamefont{{Wanas} and {Hassan}}(2014)}]{WH14}
\bibinfo{author}{\bibfnamefont{M.~I.} \bibnamefont{{Wanas}}} \bibnamefont{and}
  \bibinfo{author}{\bibfnamefont{H.~A.} \bibnamefont{{Hassan}}},
  \bibinfo{journal}{International Journal of Theoretical Physics}
  \textbf{\bibinfo{volume}{53}}, \bibinfo{pages}{3901} (\bibinfo{year}{2014}).

\bibitem[{\citenamefont{{Wanas} et~al.}(2015)\citenamefont{{Wanas}, {Osman},
  and {El-Kholy}}}]{WOR15}
\bibinfo{author}{\bibfnamefont{M.}~\bibnamefont{{Wanas}}},
  \bibinfo{author}{\bibfnamefont{S.}~\bibnamefont{{Osman}}}, \bibnamefont{and}
  \bibinfo{author}{\bibfnamefont{R.}~\bibnamefont{{El-Kholy}}},
  \bibinfo{journal}{Open Physics} \textbf{\bibinfo{volume}{13}}
  (\bibinfo{year}{2015}).

\bibitem[{\citenamefont{{Robertson}}(1932)}]{R32}
\bibinfo{author}{\bibfnamefont{H.~P.} \bibnamefont{{Robertson}}},
  \bibinfo{journal}{Ann. Math.} \textbf{\bibinfo{volume}{33}},
  \bibinfo{pages}{496} (\bibinfo{year}{1932}).

\bibitem[{\citenamefont{{Myrzakulov}}(2011)}]{M11}
\bibinfo{author}{\bibfnamefont{R.}~\bibnamefont{{Myrzakulov}}},
  \bibinfo{journal}{The European Physical Journal C}
  \textbf{\bibinfo{volume}{71}}, \bibinfo{eid}{1752} (\bibinfo{year}{2011}),
  \bibinfo{note}{arXiv: gr-qc/1006.1120}, \eprint{1006.1120}.

\bibitem[{\citenamefont{{Planck Collaboration}
  et~al.}(2014)\citenamefont{{Planck Collaboration}, {Ade}, {Aghanim},
  {Armitage-Caplan}, {Arnaud}, {Ashdown}, {Atrio-Barandela}, {Aumont},
  {Baccigalupi}, {Banday} et~al.}}]{1303.5082}
\bibinfo{author}{\bibnamefont{{Planck Collaboration}}},
  \bibinfo{author}{\bibfnamefont{P.~A.~R.} \bibnamefont{{Ade}}},
  \bibinfo{author}{\bibfnamefont{N.}~\bibnamefont{{Aghanim}}},
  \bibinfo{author}{\bibfnamefont{C.}~\bibnamefont{{Armitage-Caplan}}},
  \bibinfo{author}{\bibfnamefont{M.}~\bibnamefont{{Arnaud}}},
  \bibinfo{author}{\bibfnamefont{M.}~\bibnamefont{{Ashdown}}},
  \bibinfo{author}{\bibfnamefont{F.}~\bibnamefont{{Atrio-Barandela}}},
  \bibinfo{author}{\bibfnamefont{J.}~\bibnamefont{{Aumont}}},
  \bibinfo{author}{\bibfnamefont{C.}~\bibnamefont{{Baccigalupi}}},
  \bibinfo{author}{\bibfnamefont{A.~J.} \bibnamefont{{Banday}}},
  \bibnamefont{et~al.}, \bibinfo{journal}{Astronomy \& Astrophysics}
  \textbf{\bibinfo{volume}{571}} (\bibinfo{year}{2014}), \eprint{1303.5082}.

\bibitem[{\citenamefont{Vilenkin}(2006)}]{V2006}
\bibinfo{author}{\bibfnamefont{A.}~\bibnamefont{Vilenkin}},
  \bibinfo{journal}{Science} \textbf{\bibinfo{volume}{312}},
  \bibinfo{pages}{1148} (\bibinfo{year}{2006}).

\bibitem[{\citenamefont{{Farooq} et~al.}(2015)\citenamefont{{Farooq}, {Mania},
  and {Ratra}}}]{FMR15}
\bibinfo{author}{\bibfnamefont{O.}~\bibnamefont{{Farooq}}},
  \bibinfo{author}{\bibfnamefont{D.}~\bibnamefont{{Mania}}}, \bibnamefont{and}
  \bibinfo{author}{\bibfnamefont{B.}~\bibnamefont{{Ratra}}},
  \bibinfo{journal}{Astrophysics and Space Science}
  \textbf{\bibinfo{volume}{357}}, \bibinfo{pages}{11} (\bibinfo{year}{2015}).

\bibitem[{\citenamefont{{Dossett} and {Ishak}}(2012)}]{DI12}
\bibinfo{author}{\bibfnamefont{J.}~\bibnamefont{{Dossett}}} \bibnamefont{and}
  \bibinfo{author}{\bibfnamefont{M.}~\bibnamefont{{Ishak}}},
  \bibinfo{journal}{Phys. Rev. D} \textbf{\bibinfo{volume}{86}},
  \bibinfo{eid}{103008} (\bibinfo{year}{2012}), \eprint{1205.2422}.

\bibitem[{\citenamefont{{Zhang} et~al.}(2014)\citenamefont{{Zhang}, {Zhao},
  {Cui}, and {Zhang}}}]{ZZCZ14}
\bibinfo{author}{\bibfnamefont{J.-F.} \bibnamefont{{Zhang}}},
  \bibinfo{author}{\bibfnamefont{M.-M.} \bibnamefont{{Zhao}}},
  \bibinfo{author}{\bibfnamefont{J.-L.} \bibnamefont{{Cui}}}, \bibnamefont{and}
  \bibinfo{author}{\bibfnamefont{X.}~\bibnamefont{{Zhang}}},
  \bibinfo{journal}{European Physical Journal C} \textbf{\bibinfo{volume}{74}},
  \bibinfo{pages}{3178} (\bibinfo{year}{2014}), \eprint{1409.6078}.

\bibitem[{\citenamefont{{Xie} and {Shirafuji}}(1996)}]{XSH96}
\bibinfo{author}{\bibfnamefont{H.-j.} \bibnamefont{{Xie}}} \bibnamefont{and}
  \bibinfo{author}{\bibfnamefont{T.}~\bibnamefont{{Shirafuji}}}
  (\bibinfo{year}{1996}), \eprint{9603006}.

\bibitem[{\citenamefont{{Hojman} et~al.}(1978)\citenamefont{{Hojman},
  {Rosenbaum}, {Ryan}, and {Shepley}}}]{HRR78}
\bibinfo{author}{\bibfnamefont{S.}~\bibnamefont{{Hojman}}},
  \bibinfo{author}{\bibfnamefont{M.}~\bibnamefont{{Rosenbaum}}},
  \bibinfo{author}{\bibfnamefont{M.~P.} \bibnamefont{{Ryan}}},
  \bibnamefont{and} \bibinfo{author}{\bibfnamefont{L.~C.}
  \bibnamefont{{Shepley}}}, \bibinfo{journal}{Phys. Rev. D}
  \textbf{\bibinfo{volume}{17}}, \bibinfo{pages}{3141} (\bibinfo{year}{1978}).

\bibitem[{\citenamefont{{Mukku} and {Sayed}}(1979)}]{Mukku:1979}
\bibinfo{author}{\bibfnamefont{C.}~\bibnamefont{{Mukku}}} \bibnamefont{and}
  \bibinfo{author}{\bibfnamefont{W.~A.} \bibnamefont{{Sayed}}},
  \bibinfo{journal}{Physics Letters B} \textbf{\bibinfo{volume}{82}},
  \bibinfo{pages}{382} (\bibinfo{year}{1979}).

\bibitem[{\citenamefont{{Rochev}}(1974)}]{R74}
\bibinfo{author}{\bibfnamefont{V.~E.} \bibnamefont{{Rochev}}},
  \bibinfo{journal}{Theoretical and Mathematical Physics}
  \textbf{\bibinfo{volume}{18}}, \bibinfo{pages}{160} (\bibinfo{year}{1974}).

\bibitem[{\citenamefont{{Hammond}}(1990)}]{H90}
\bibinfo{author}{\bibfnamefont{R.~T.} \bibnamefont{{Hammond}}},
  \bibinfo{journal}{Classical and Quantum Gravity}
  \textbf{\bibinfo{volume}{7}}, \bibinfo{pages}{2107} (\bibinfo{year}{1990}).

\bibitem[{\citenamefont{{Hehl} and {Obukhov}}(2001)}]{HO01}
\bibinfo{author}{\bibfnamefont{F.~W.} \bibnamefont{{Hehl}}} \bibnamefont{and}
  \bibinfo{author}{\bibfnamefont{Y.~N.} \bibnamefont{{Obukhov}}}, in
  \emph{\bibinfo{booktitle}{Gyros, Clocks, Interferometers ...: Testing
  Relativistic Gravity in Space}}, edited by
  \bibinfo{editor}{\bibfnamefont{C.}~\bibnamefont{{L{\"a}mmerzahl}}},
  \bibinfo{editor}{\bibfnamefont{C.~W.~F.} \bibnamefont{{Everitt}}},
  \bibnamefont{and} \bibinfo{editor}{\bibfnamefont{F.~W.} \bibnamefont{{Hehl}}}
  (\bibinfo{year}{2001}), vol. \bibinfo{volume}{562} of
  \emph{\bibinfo{series}{Lecture Notes in Physics, Berlin Springer Verlag}}, p.
  \bibinfo{pages}{479}, \eprint{gr-qc/0001010}.

\end{thebibliography}

\end{document}